\documentclass[twocolumn, tighten]{aastex62}

\accepted{7 December 2018, ApJL in press.}

\shorttitle{High definition study of the HD~163296 disk}
\shortauthors{Isella et al.}

\begin{document}

\title{The Disk Substructures at High Angular Resolution Project (DSHARP) - IX\\
A high definition study of the HD~163296 planet forming disk.}

\correspondingauthor{Andrea Isella}
\email{isella@rice.edu}

\author[0000-0002-0786-7307]{Andrea Isella}
\affiliation{Department of Physics and Astronomy, Rice University 6100 Main Street, MS-108, Houston, TX 77005, USA}

\author[0000-0001-6947-6072]{Jane Huang}
\affiliation{Harvard-Smithsonian Center for Astrophysics, 60 Garden Street, Cambridge, MA 02138, USA}

\author{Sean M. Andrews}
\affiliation{Harvard-Smithsonian Center for Astrophysics, 60 Garden Street, Cambridge, MA 02138, USA}

\author[0000-0002-7078-5910]{Cornelis P. Dullemond}
\affiliation{Zentrum für Astronomie, Heidelberg University, Albert Ueberle Str. 2, 69120 Heidelberg, Germany}

\author[0000-0002-1899-8783]{Tilman Birnstiel}
\affiliation{University Observatory, Faculty of Physics, Ludwig-Maximilians-Universit\"at M\"unchen, Scheinerstr. 1, 81679 Munich, Germany}

\author[0000-0002-8537-9114]{Shangjia Zhang}
\affiliation{Department of Physics and Astronomy, University of Nevada, Las Vegas, 4505 S. Maryland Pkwy, Las Vegas, NV, 89154, USA}

\author[0000-0003-3616-6822]{Zhaohuan Zhu}
\affiliation{Department of Physics and Astronomy, University of Nevada, Las Vegas, 4505 S. Maryland Pkwy, Las Vegas, NV, 89154, USA}

\author{Viviana V.~Guzm{\'a}n}
\affiliation{Joint ALMA Observatory, Avenida Alonso de C{\'o}rdova 3107, Vitacura, Santiago, Chile}
\affiliation{Instituto de Astrof{\'i}sica, Pontificia Universidad Cat{\'o}lica de Chile, Av.~Vicu{\~n}a Mackenna 4860, 7820436 Macul, Santiago, Chile}

\author{Laura M.~P{\'e}rez}
\affiliation{Departamento de Astronom{\'i}a, Universidad de Chile, Camino El Observatorio 1515, Las Condes, Santiago, Chile}

\author{Xue-Ning Bai}
\affiliation{Institute for Advanced Study and Tsinghua Center for Astrophysics, Tsinghua University, Beijing 100084, China}

\author[0000-0002-7695-7605]{Myriam Benisty}
\affiliation{Unidad Mixta Internacional Franco-Chilena de Astronom\'{i}a, CNRS/INSU UMI 3386, Departamento de Astronom\'ia, Universidad de Chile, Camino El Observatorio 1515, Las Condes, Santiago, Chile}
\affiliation{Univ. Grenoble Alpes, CNRS, IPAG, 38000 Grenoble, France}

\author{John M.~Carpenter}
\affiliation{Joint ALMA Observatory, Avenida Alonso de C{\'o}rdova 3107, Vitacura, Santiago, Chile}

\author{Luca Ricci}
\affiliation{Department of Physics and Astronomy, California State University Northridge, 18111 Nordhoff Street, Northridge, CA 91130, USA}

\author{David J.~Wilner}
\affiliation{Harvard-Smithsonian Center for Astrophysics, 60 Garden Street, Cambridge, MA 02138, USA}







\begin{abstract}
ALMA observations of protoplanetary disks acquired by the Disk Substructure at High Angular Resolution Project (DSHARP) resolve the dust and gas emission on angular scales as small as 3 astronomical units, offering an unprecedented detailed view  of the environment where planets form. In this article, we present and discuss observations of the HD~163296 protoplanetary disk that imaged the 1.25~mm dust continuum and $^{12}$CO J=2-1 rotational line emission at a spatial resolution of 4 and 10 au, respectively. The continuum observations resolve and allow us to characterize the previously discovered dust rings at radii of 67 and 100 au. They also reveal new small scale structures, such as a dark gap at 10 au, a bright ring at 15 au, a dust crescent at a radius of 55 au, and several fainter azimuthal asymmetries. The observations of the CO and dust emission inform about the vertical structure of the disk and allow us to directly constrain the dust extinction optical depth at the dust rings. Furthermore, the observed asymmetries in the dust continuum emission corroborate to the hypothesis that the complex structure of the HD~163296 disk is the result of the gravitational interaction with yet unseen planets. 
\end{abstract}



\section{Introduction} 
\label{sec:intro}

In recent years, millimeter-wave interferometers and near infrared high contrast cameras have imaged nearby protoplanetary systems at unprecendented angular resolution in both continuum and line emission \citep[e.g., ][]{andrews16, brogan15, isella16, keppler18, sallum15}. Although still limited to a few objects, these observations inform about the processes responsible for the formation of planets. 

Several circumstellar disks observed at a spatial resolution better than 50 au reveal ring-like features in the emission of small and large dust particles, as well as of the molecular gas \citep{andrews11, andrews11b, boehler18, dong17, dong18, fedele17, fedele18, isella13, isella14, tang17, vandermarel15, vandermarel16, vandermarel18, zhang14, zhang16}. In particular, the homogeneous survey performed by the Disk Substructure at High Angular Resolution Project (DSHARP) has revealed that multiple-ring  systems are ubiquitous among the most massive circumstellar disks (Andrews et al. 2018). The number of rings and their structure vary substantially from object to object, and in some cases, even within the same disk (Huang et al. 2018).

A myriad of theoretical models have been proposed to explain the formation of rings in circumstellar disks. These include the interaction between the disk and yet-unseen giant planets \citep[e.g.][]{bryden99, zhu14, dong18}, sharp opacity variations at gas-solid phase transitions \citep{banzatti15, zhang15, okuzumi16}, dust accumulations at the edge of low viscosity regions \citep{flock15, miranda17}, and zonal flows via spontaneous concentration of net vertical flux \citep{bai14b, bethune17, suriano18}. However, to date, planet-disk interaction models have been the most successful in explaining the observed structures \citep{jin17, liu18, dong18}. Taken at face value, these results tend to suggest the existence of a population of young gas giant planets orbiting at several tens of au from the central star which challenge current planet formation models. Furthermore, the link between planets and rings is supported by the direct detection of possible young planets in the PDS~70 and LkCa~15 systems \citep{keppler18, sallum15}, although the nature of some candidates is debated \citep{mendigutia18, thalmann16}.

In this paper, we present new ALMA observations of the 1.25~mm dust continuum and $^{12}$CO J=2-1 line emission of the circumstellar disk around the Herbig Ae star HD~163296. The observations achieve a spatial resolution of 4 au and deliver the sharpest images of this source obtained to date. At a distance of 101 pc \citep{gaia_dr2}, HD~163296 is in several respects the poster child of disks thought to be perturbed by planets. The star is surrounded by a huge (more than 1000 au in diameter) Keplerian disk whose mass has been estimated to range between 0.01 and 0.15 M$_\odot$ \citep{isella07,muro18,tilling12}. Previous ALMA observations that resolved the gas and dust emission on spatial scales of 20 au, revealed the presence of three circular gaps in the disk density with radii of 45, 87, and 140 au, and, correspondingly, three dense rings centered at 68, 100, and 160 au  \citep{isella16,zhang16}. The recent detection of deviations from Keplerian rotation at the location of the gaps and rings by \cite{teague18} confirms that these structures correspond to variations in the gas pressure (i.e., variations in the gas density and/or temperature), as opposed to radial changes in the dust opacity. The comparison between the observations and planet-disk interaction models indicates that the density gaps might have been carved by planets with masses between 0.5 and 1 M$_J$ orbiting at 48, 86, and 131 au from the central star \citep[][]{liu18, vandermarel18b}. However, models that assume very low viscosity ($\alpha < 10^{-4}$) suggest that the observed multiple ring structure may also result from the gravitational interaction with a single planet less massive than Saturn at 100 au orbital radius \citep{dong18}. Furthermore, the presence of an additional planet has been proposed based on the detection of local deviations from Keplerian velocity. Current models suggest that this planet might be located at 260 au from the star and has a mass of about 2 M$_J$ \citep{pinte18}. 

The HD~163296 system has also been  the target of a number of  optical and infrared high-contrast imaging campaigns aimed at characterizing the morphology of the disk in scattered light emission and detecting low mass companions \citep{benisty10, garufi14, grady00, muro18, wisniewski08}. Whereas these observations revealed rings similar to those observed at millimeter wavelengths, they did not detect any stellar or planetary mass object at orbital radii larger than 25 au and down to a mass sensitivity of a few M$_J$ \citep{guidi18}. Thanks to the unprecedented angular resolution, our new ALMA observations provide detailed information on the morphology of the HD~163296 disk that help understanding the origin of the observed structures. 

The structure of the paper is the following. The observations and data reduction are discussed in Section~\ref{sec:data}. The map of the dust continuum emission and its analysis are presented in Section~\ref{sec:cont}, while CO data are discussed in Section~\ref{sec:co}. The implications of our results in the context of constraining the temperature of the disk and the origin of the observed structures are presented in Section~\ref{sec:discussion}. Finally, the main results of our investigation are summarized in Section~\ref{sec:conc}.

\section{Observations} 
\label{sec:data}

\begin{figure*}[t]
\includegraphics[width=\textwidth]{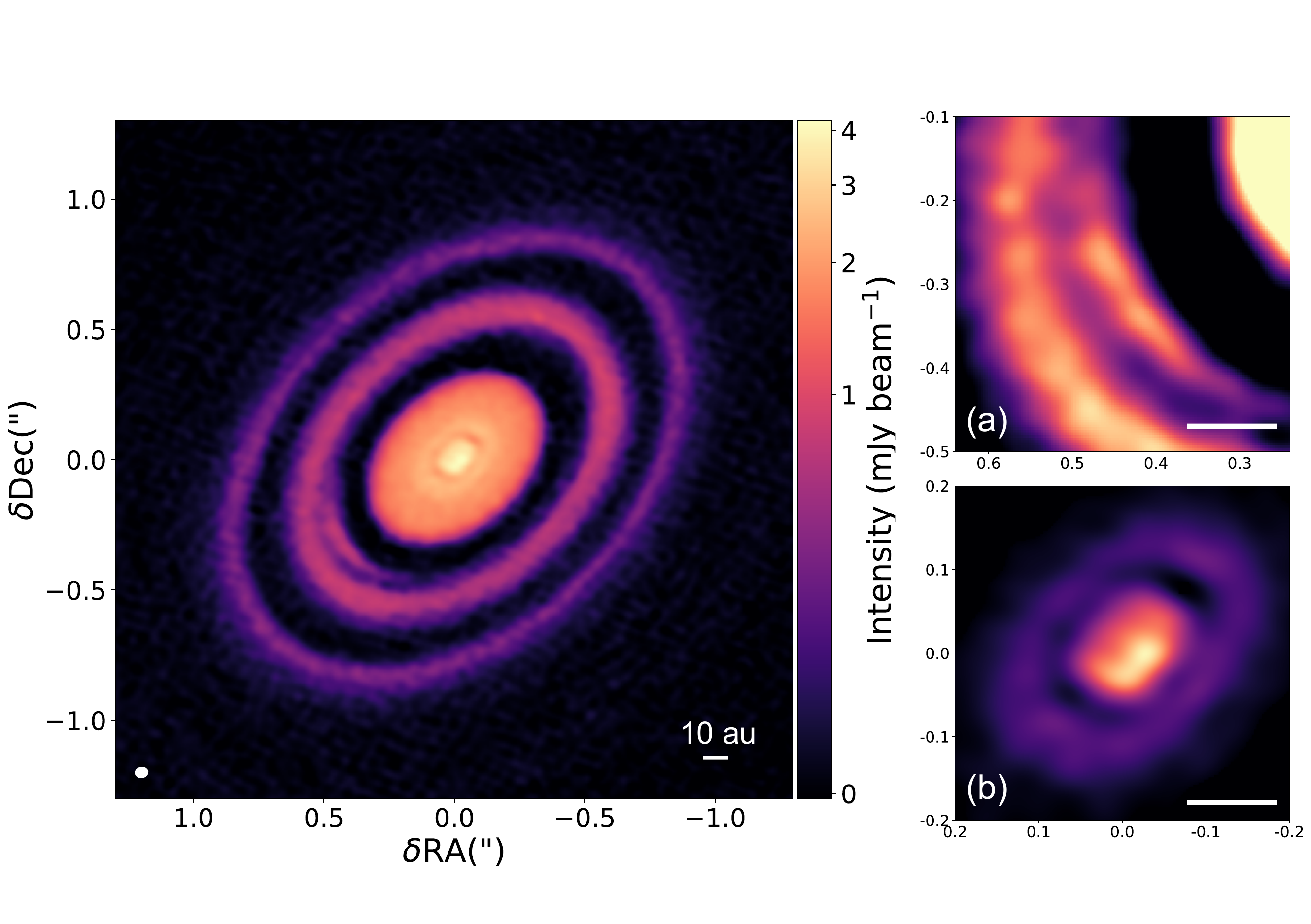}
\caption{Map of the HD~163296 disk recorded in the 1.25 mm continuum. The angular resolution of the observations is 0.038\arcsec$\times$0.048\arcsec and is indicated by the white ellipse in the bottom left corner of the left panel. At the source distance of 101 pc, the spatial resolution is 3.8au$\times$4.8au.  Inset (a) and (b) show zoom-in view of two asymmetric features revealed by the observations. The rms noise is 0.023 mJy beam$^{-1}$.  The horizontal segments indicate a spatial scale of 10 au.\label{fig:cont}}
\end{figure*}

The observations used in this study were obtained during the
DSHARP ALMA Large Program (2016.1.00484.L) and are discussed in 
detail in Andrews et al. (2018). In brief, the HD~163296 disk was 
observed in Band 6 ($\lambda \sim 1.3$ mm) in September 2017 in 
configuration C40-8, which delivered baselines $B$ between 41 m and 
5.8 km, and a theoretical angular resolution $\lambda/B_{max}\sim 0.04\arcsec$.  
Archival data from project 2013.1.00366.S \citep{flaherty15} and 
project 2013.1.00601.S \citep{isella16} were used to improve 
sensitivity and uv-coverage on shorter baselines.

An initial calibration of each dataset was produced by the ALMA pipeline.
A visual inspection of the data obtained from project 2013.1.00366.S revealed 
a number of visibilities characterized by very noisy amplitudes that required manual 
flagging.  Short baseline observations were independently self-calibrated 
and re-centered to account for shifts in the position of the disk 
due to the proper motion of the source. We compared the flux calibration of each 
track and adjusted the flux scale so that the source visibility amplitude measured on  
overlapping baselines agrees within 5\%.
Finally, we combined all the tracks and performed phase and 
amplitude self-calibration on both short and long baseline data. 
The self-calibration procedure resulted in an improvement of 43\%
in the peak signal-to-noise ratio in the dust continuum image. 
A complete description of the DSHARP data calibration procedure is presented in 
Andrews et al. (2018), while the  CASA script used to calibrate and image HD~163296 data, including 
the manual flagging and flux rescaling,  is available 
online at \url{https://almascience.org/alma-data/lp/DSHARP}.
The 1.25 mm dust continuum emission was imaged using the CASA task {\it tclean} 
and a robust parameter of -0.5 resulting in a synthesized beam FWHM of 0.038\arcsec$\times$0.048\arcsec,  corresponding to  
a spatial resolution of 3.8$\times$4.8 au at the distance of the source. The  rms noise 
is 0.023 mJy beam$^{-1}$,  and the peak signal to noise ratio is 185.

The complex gain solutions of the self-calibration of the continuum emission were then applied to the $^{12}$CO data. 
The line was observed at the same angular resolution of the continuum ($\sim$0.04\arcsec), but we imaged it using Briggs robust=0.5 (no uv-tapering) 
to achieve higher signal-to-noise ratio (see Andrews et al. 2018 for more details about CO imaging).
The FWHM of the synthesized beam of the presented CO maps is 0.104\arcsec$\times$0.095\arcsec, which is about half the beam of the CO observations published in \cite{isella16}, and about 7 times smaller than the resolution of the ALMA science verification data \citep{rosenfeld13a, degregorio-monsalvo13}. Channels are spaced in velocity by 0.32 km s$^{-1}$, but, due to Hanning smoothing, the velocity resolution is 0.64 km s$^{-1}$.  The velocity grid for HD~163296 slightly differs from the fiducial channel spacing of 0.35 km s$^{-1}$ adopted for the other DSHARP sources. The  rms noise per channel is 0.84 mJy beam$^{-1}$.

\section{Dust continuum emission}
\label{sec:cont}

The map of the 1.25 mm continuum emission (Figure~\ref{fig:cont}) features two bright elliptical rings previously reported by \cite{isella16}, as well as three new  morphological features: an arc of emission inside the first ring (inset a), an inner dark gap and bright  ring at about 10 au and 15 au, respectively, and a central azimuthal asymmetry (inset b). In this section we discuss the morphology of the dust rings, constrain the radial profile of the continuum emission, and investigate the presence of asymmetric structures in the distribution of the solid material. 

\subsection{Morphology of the dust rings}  
\label{sec:cont_morph}

As a first step toward characterizing the structure of the dust rings, we fit the crests and troughs of the continuusm emission with circles defined by the position of the center ($\Delta x_0$, $\Delta y_0$), the radius $r$, the inclination $i$ at which they are observed, and the position angle (PA) of their projected major axis. The fit is performed as discussed in Huang et al. (2018). In brief, we locate the points (x,y) corresponding to the radial maxima (or minima) of each ring (or gap), and use the python {\it emcee} package to calculate the ellipses that best reproduce this set of points. The best fit parameters are listed in Table \ref{table:rings}. Bright rings and dark gaps  are labeled with the suffixes B and D, respectively, while the number appearing in the feature name corresponds to their radius in au. The uncertainties on the best fit values correspond to the 16th and 84th percentile of the marginalized posterior probability distribution.

\begin{deluxetable*}{lcccccc}
\tabletypesize{\scriptsize}
\tablecaption{Results of elliptical fitting to the bright and dark rings observed in the 1.25~mm continuum map. \label{table:rings}}
\tablehead{
\colhead{Feature} \vspace{-0.2cm}& \colhead{$\Delta x_0$ (mas)} & \colhead{$\Delta y_0$(mas)} & \colhead{$r$(")} & \colhead{$r$(au)} & \colhead{$i$(\arcdeg)} & \colhead{PA(\arcdeg)} 
}
\colnumbers
\startdata
D10  & -6.2$\pm$0.8 &  10.5$\pm$0.7  & 0.0987$\pm$0.0009 & 9.96$\pm$0.07  & 37.9$\pm$1.2  & 127.90$\pm$1.9 \\
B14  & -4.8$\pm$0.7  &  11.4$\pm$0.7  & 0.1430$\pm$0.0010 & 14.44$\pm$0.07  & 47.24$\pm$0.64  & 131.10$\pm$0.88 \\ 
D45  & -5.4$\pm$2.0  &  5.2$\pm$1.9  & 0.4433$\pm$0.0026 & 44.77$\pm$0.19  & 42.22$\pm$0.67  & 133.67$\pm$0.95 \\
B67  & -5.3$\pm$1.0  &  7.3$\pm$1.1  & 0.6633$\pm$0.0014 & 66.99$\pm$0.11  & 46.78$\pm$0.21  & 133.13$\pm$0.29 \\ 
D86  & 6.7$\pm$2.3  &  -0.3$\pm$2.3  & 0.8575$\pm$0.0028 & 86.61$\pm$0.22  & 47.34$\pm$0.32  & 132.78$\pm$0.45 \\
B100 & -2.3$\pm$0.9  &  8.6$\pm$0.9  & 0.9870$\pm$0.0011 & 99.69$\pm$0.08  & 46.59$\pm$0.11  & 133.46$\pm$0.15 \\
D141 & -0.2$\pm$9.9  & 6.4$\pm$9.5  & 1.3923$\pm$0.0126   & 140.62$\pm$0.96 & 47.2$\pm$0.9  & 131.3$\pm$1.2 \\
B159 & -16.2$\pm$12.0 &  1.1$\pm$12.0 & 1.572$\pm$0.016 & 158.7$\pm$1.2   & 45.7$\pm$1.0  & 132.0$\pm$1.4 \\
\hline
\hline
mean* & -3.5$\pm$0.5 &  9.1$\pm$0.5 & - & - & 46.7$\pm$0.1 & 133.3$\pm$0.1
\enddata
\tablecomments{*the parameters of D10 and D45 gaps were not used to calculate the mean values.}
\end{deluxetable*}

The outermost features B67, D86, B100, D145, and B155,  are well described by concentric circles with an average inclination of (46.7$\pm$0.1)\arcdeg\, and average position angle of (133.3$\pm$0.1)\arcdeg. Taken at face value, the circular model fitting indicates that D10 and D45 have lower inclinations. We believe that the lower inclination of D45 is due to the  presence of the dust crescent centered at 55 au, which partially fills the gap making it appear more circular (i.e. less inclined), while the lower inclination of D10 is compatible with the effect of beam smearing which is discussed in more details below.

\begin{figure*}[t]
\includegraphics[width=0.430\linewidth]{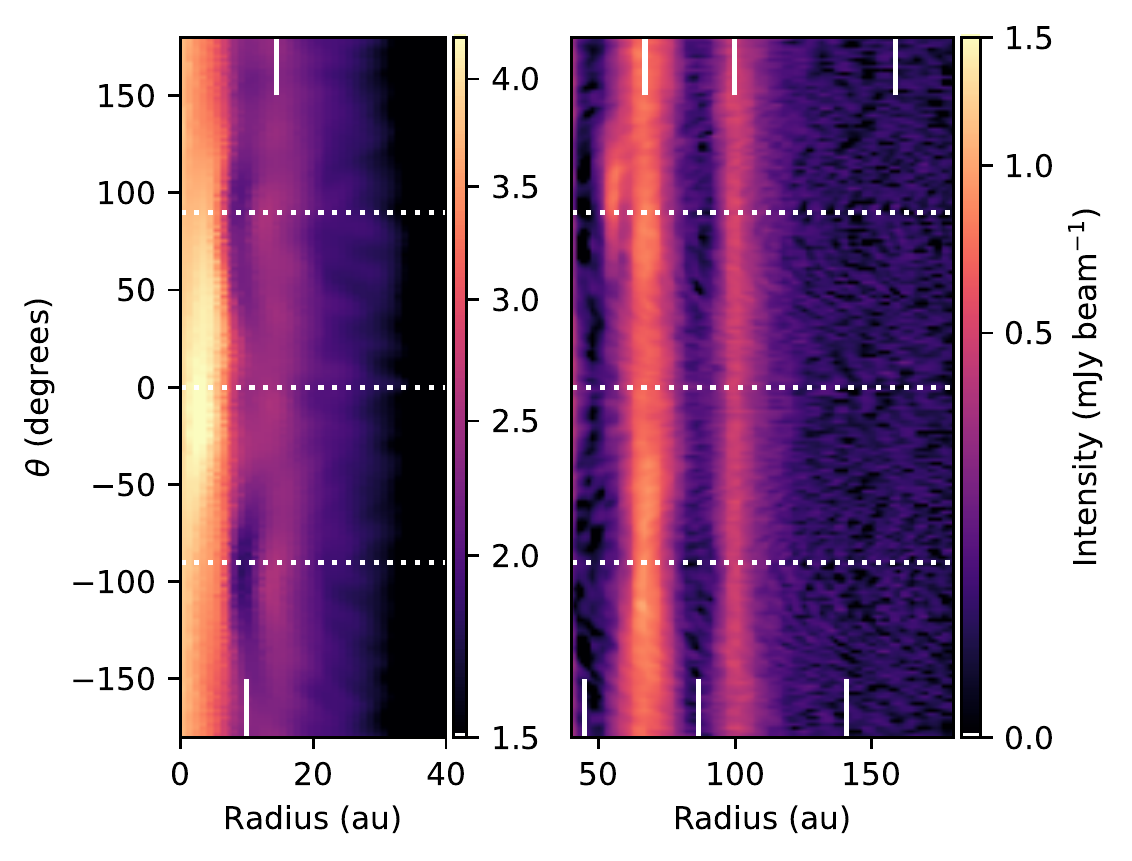}
\includegraphics[width=0.569\linewidth]{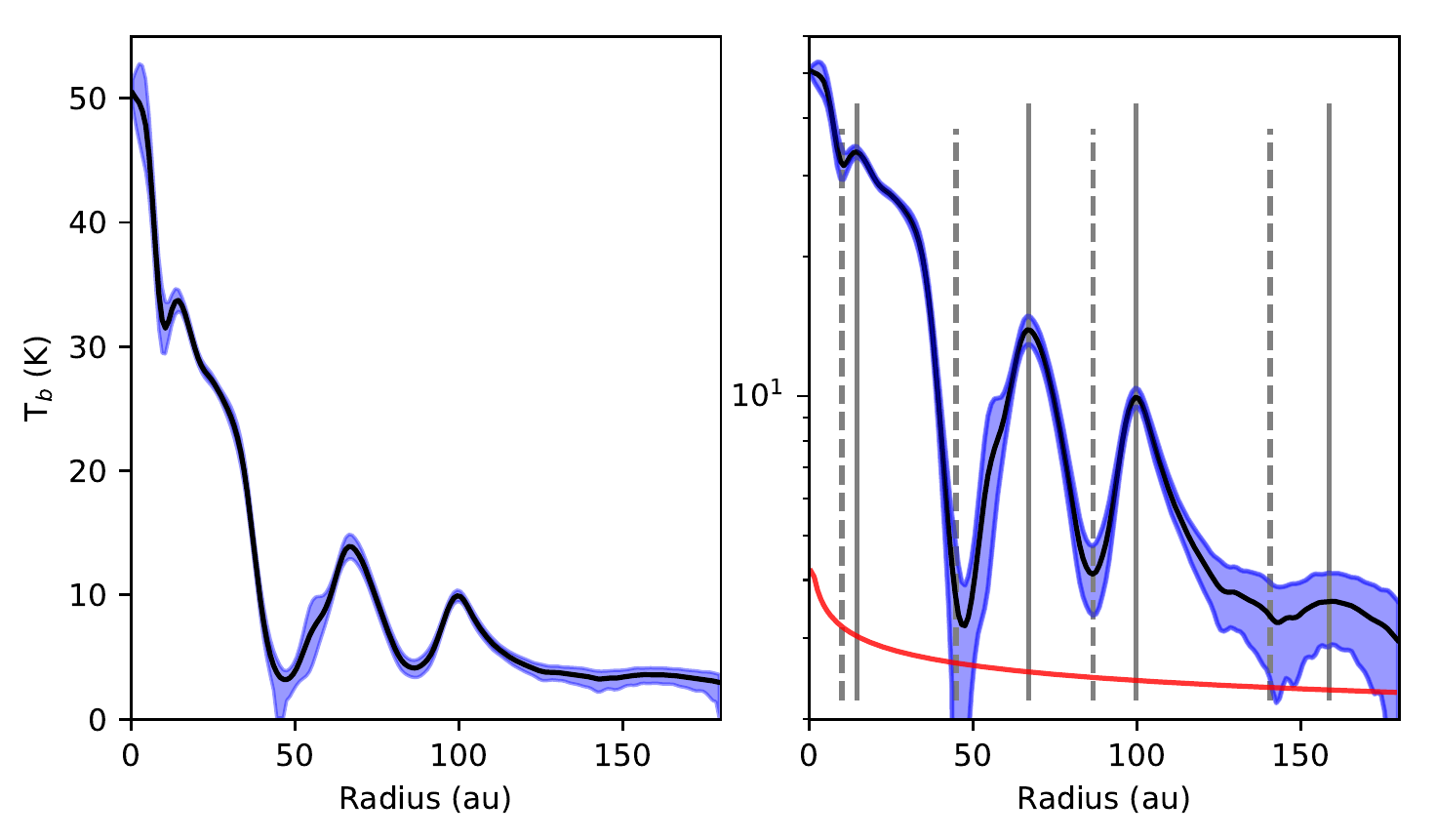}
\caption{\label{fig:cont_deproj} Left: Polar map of the 1.25~mm dust continuum emission deprojected using a disk inclination of 46.7\arcdeg. Two different color scales are used to highlight the dust intensity structures observed inside and outside of 40 au. The angle $\theta$ indicates the elongation from the disk apparent minor axis measures positive East of North. The top and bottom white bars indicate the position of the bright rings and dark gaps, respectively. 
Center: Azimuthal average of dust continuum emission expressed in units of brightness temperature $T_b^{pl}$. Right: Same as in the center but plotted on a logarithmic temperature scale. The blue curves indicate the dispersion around the mean value, while the red curve indicates the temperature corresponding to 3$\times$ the rms noise of the azimuthally averaged intensity calculated taking into consideration both the rms noise of the continuum map and the number of resolution elements in each radial bin. Vertical dashed lines indicate the position of the dark rings, while vertical solid lines indicate the position of bright rings.   }
\end{figure*}

Figure~\ref{fig:cont_deproj} shows the deprojected map and the azimuthally averaged radial profile of the continuum emission obtained using the values for the disk inclination and position angle derived from the outermost rings. In the deprojected map, the outer rings appear as vertical stripes, confirming that these structures have indeed similar inclination and position angle, and that they are intrinsically circular in shape. The intensity ratio between B14 and D10 varies with the 
azimuthal angle reaching a maximum of about 1.2 along the disk major axis ($\theta=\pm90$\arcdeg) and a minimum of about 1 along the disk minor axis 
($\theta=0$\arcdeg). This variation can be explained with the fact that the effective spatial resolution for an inclined disk is higher along the 
disk major axis compared to the disk minor axis.  In practice, B14 and D10 are spatially resolved only along the disk major axis, while they blend 
together along the disk minor axis.

Azimuthal variations in the contrast between bright rings and dark gaps are also visible for the outer pairs. For example, along the westward semi-major axis ($\theta=-90$\arcdeg), the intensity reaches a minimum consistent with the noise of the observations at the center of D45, and a maximum of 0.91 mJy beam$^{-1}$ at the center of the adjacent bright ring B67. This corresponds to an intensity ratio larger than 37. In the opposite direction, i.e. along the eastward disk semi-major axis ($\theta=90$\arcdeg), the intensity ratio B67/D45 is 13. As comparison, the intensity ratio along the southward ($\theta=180$\arcdeg) and northward ($\theta=0$\arcdeg) direction of the disk minor axis is about 7 and 5, respectively (see Table~\ref{table:int_ratio} for a summary of the intensity ratio among rings pairs). 
The difference between intensity ratios measured along the disk major and minor axes is likely due to beam smearing, while the difference between the intensity ratios measured along the disk major axis suggests the presence of azimuthal asymmetries in the dust emission.

\begin{deluxetable}{lccccc}
\tabletypesize{\scriptsize}
\tablecaption{Intensity ratios between bright rings and dark gaps along different radial directions \label{table:int_ratio}}
\tablehead{
\colhead{Pair} \vspace{-0.3cm} & \colhead{+90\arcdeg} & \colhead{-90\arcdeg} & \colhead{0\arcdeg} & \colhead{180\arcdeg}& \colhead{Mean}\\ \vspace{-0.2cm}
 & \colhead{(Maj E)} & \colhead{(Maj W)} & \colhead{(Min N)} & \colhead{(Min S)} &}
\colnumbers
\startdata
B14/D10 & 1.2 & 1.1 & 1.0 & 0.9 & 1.1\\
B67/D45 & 13.3 & $>$37 & 6.6 & 4.7 & $>$17\\
B100/D86 & 11.7 & 7.1 & 4.5 & 2.5 & 7.0\\
B159/D141 & 2.1 & 1.1 & 1.0 & 0.5 & 1.3\\
\enddata
\tablecomments{(1) Name of the pair. (2 and 3) Intensity ratio along the apparent disk major axis in the eastward ($\theta=+90\arcdeg$) and westward ($\theta=-90\arcdeg$) directions, respectively. (4 and 5) Intensity ratio along the apparent disk minor axis in the northward ($\theta=0\arcdeg$) and southward ($\theta=-180\arcdeg$) directions, respectively. (6) Intensity ratio of the azimuthally averaged intensity profile}
\end{deluxetable}

The center and right panels of Figure~\ref{fig:cont_deproj} show the azimuthally averaged profile of the deproject continuum emission expressed in units of brightness temperature $T^{pl}_{b}$ calculated using the full Planck equation as 
\begin{equation}
\label{eq:tbright}
    T^{pl}_b = \frac{h \nu}{k_b} \left[ \ln \left( \frac{2 h \nu^3 \Omega}{c^2 F_\nu}  +1 \right) \right]^{-1},
\end{equation}
where the $F_\nu$ is the flux density integrated over a beam defined by the solid angle $\Omega=\pi \theta_{min}\theta_{max}/(4\ln2)$, where $\theta_{min}$ and $\theta_{max}$ are the minimum and maximum FWHM of the synthesized clean beam. The brightness temperature varies between about 50~K (corresponding to an intensity of 3.8 mJy beam$^{-1}$) in the innermost disk regions down to about 3 K (or a flux density of 0.0025 mJy beam$^{-1}$) at about 180 au. The pairs B67/D45 and B100/D86 have brightness temperature ratios of 4.3 and 2.4, respectively. Note that the difference between brightness temperature and intensity ratios is due to the non-linearity of Equation~\ref{eq:tbright}.

\subsection{Width of the continuum rings} 
\label{cont:width}

\begin{deluxetable*}{lcccccc}
\tabletypesize{\scriptsize}
\tablecaption{Results of Gaussian model fitting of the dust continuum emission\label{table:gaussfit}}
\tablehead{\\
 \vspace{-0.1cm} & \multicolumn{3}{c}{------ Radial intensity profile ------} &  \multicolumn{3}{c}{ ------ Visibilities ------} \\
\colhead{Feature} \vspace{-0.2cm} &  \colhead{$r$ (au)} & \colhead{$w_d$ (au)} & \colhead{I (Jy/as$^2$)} &  \colhead{$r$ (au)} & \colhead{$w_d$ (au)} & \colhead{I (Jy/as$^2$)}
}
\colnumbers
\startdata
-     	&  - 	 	&  -	    	&  - 		& 3.8$\pm$0.1 			        & 2.3$\pm$0.1		& $1.7\pm0.2$  \\
B14  & - 	 	& -		&	-  	& $15.5\pm0.2$			& $8.7\pm0.2$		& 1.1$\pm$0.1  \\
-     	&  - 	 	&- 		& - 		& 31.8$\pm$0.3 			& 4.4$\pm$0.1 		& 0.62$\pm$0.06\\
B67  & 67.7   	& 6.84	& 0.38 	& 67.08$\pm$0.03		 	& 6.56$\pm$0.05		& 0.38$\pm$0.02\\
B100& 100.0 	& 4.67 	& 0.24  	& 101.16$\pm$0.04 			& 5.8$\pm$0.1 			& 0.20$\pm$0.01\\
\enddata
\tablecomments{Column 2 to 4 list the best fit parameters for the radius, width, and intensity, respectively, of the Gaussian components used to model the azimuthally averaged profile of the dust continuum emission as in Dullemond et al. (2018).  Column 5 to 7 list the same quantities obtained by modeling the observations in the Fourier space. Additional parameters not reported in the table are the disk inclination $i=46.7\arcdeg \pm 0.1\arcdeg$, the disk position angle PA$=132.8\arcdeg \pm 0.1\arcdeg$, and the offset of the center of the emission relative to the phase center of the observations $\delta x_0=-8.7\pm0.1$ mas, $\delta y_0=-3.0\pm0.1$ mas.}
\end{deluxetable*}

Dullemond et al. (2018) finds that many of the continuum rings revealed by DSHARP observations, including the rings B67 and B100 of the HD~163296 disk, have a radial width narrower than the estimated pressure scale height of the gaseous disk, and conclude  that this is strong evidence that these rings have formed by radial migration of dust particles toward local maxima of the gas pressure. Dullemond et al. (2018) defines the width of a ring as the dispersion of a Gaussian function that reproduces the deprojected and azimuthally averaged radial profile of the dust emission across the ring itself. This modeling approach is very fast,  but it does not account for the observational noise, nor for the discrete sampling of the uv plane performed by an interferometer, or for the fact that the synthesized beam is not circular. 

Here,  we compare the widths of the HD~163296 rings measured by Dullemond et al. (2018) with those estimated by a more accurate, but much slower, modeling of the continuum visibilities in the uv domain.  We  start by assuming that the dust emission is axisymmetric and expressed by a sum of $n$ Gaussian functions
\begin{equation}
\label{eq:gauss}
I(r) = \sum_{i=1}^n{I_i e^{-(r-r_i)^2/2 w_{d,i}^2}}.  
\end{equation}
We generate synthetic 2D images of the continuum emission that are inclined and rotated to simulate all possible disk viewing angles.
The center of emission is allowed to vary with respect to the center of the image (i.e. the phase center). Synthetic visibilities ($V^{mod}$) are then calculated by taking the Fourier transform of the image at the spatial frequencies of the observed visibilities ($V^{obs}$).  The goodness of the model fit is evaluated through a usual $\chi^2$ test, where  $\chi^2 =\sum{w(V^{obs}-V^{mod})^2}$, and $w$ is the weight of each visibility measurement provided by the ALMA pipeline. The calculation of $\chi^2$ was performed using the cpu-based version of python package {\it galario} \citep{tazzari17}. Finally, the $\chi^2$ is used to sample the posterior likelihood using the python package {\it emcee} \citep{foreman-mackey13}.  

An initial exploration of the parameter space indicates that  a minimum of 5 Gaussian components are required to reproduce the observed continuum emission. In total, the model therefore has 19 free parameters: the coordinates ($x_0$, $y_0$) of the center of the emission, the disk inclination $i$ and position angle PA, and the radius $r$, peak intensity $I$, and width $w_d$ of five Gaussian rings.  We run {\it emcee} using 200 walkers initialized around the parameter values obtained by a Gaussian fitting of the radial intensity profile as in Dullemond et al. (2018). Each walker is evolved for an initial burn-in run of $10^4$ steps, after which they all converge toward a stable set of parameters. Following \cite{goodman10}, we sample the posterior distribution by letting {\it emcee} run until convergence, which is established based on the mean autocorrelation time of all the model parameters.  We find that convergence is achieved after about $4\times10^4$ steps. The optimal model parameters are then calculated as the median of the marginalized probability distribution and are listed in Table~\ref{table:gaussfit}, while the related uncertainties  correspond to the 0.3th and 99.7th percentile of the marginalized posterior probability distribution. This would be equivalent to a 3$\sigma$ uncertainty for a Gaussian posterior distribution.  An image constructed from the residual visibilities is shown in Figure~\ref{fig:fitvis}. 

For B67, we find a width of  $\sim 6.6$~au, that is very close to the 6.8 au derived by Dullemond et al. (2018). Conversely, the width of B100  is about 25\% 
larger than that measured by fitting the intensity profile.  As noted in Figure 3 of Dullemond et al. (2018), the intensity radial profile of B100 has a peaked central region 
and bright wings that are not well reproduced by a Gaussian function.  The value of the width in Dullemond et al. (2018) results from fitting only the 
innermost part of the ring between 94 au and 104 au, while our value resulted from modeling the global profile including the bright wings of the ring. 
The fact that the two modeling procedures achieve the same result for rings that are intrinsically Gaussian (i.e., B67) indicates that the results of 
Dullemond et al. (2018) are independent of the exact shape of the beam, or, more generally, from non linear effects caused by the discrete uv-sampling 
and image deconvolution. In practice, the uv-sampling of the observation is sufficiently uniform and the signal-to-noise ratio of the continuum map is sufficiently high 
to allow to perform the data analysis in the image domain without loss of accuracy. This is a great advantage since modeling 
the dust rings in the image domain takes only a few seconds, compared to the several days, or weeks, required to perform the model fitting in the Fourier domain. 

Our analysis indicates that the ring B14 has a radius of about 15.5 au, slightly larger than the value calculated by fitting the crest of the emission as discussed in Section~\ref{sec:cont_morph} (see also Table~\ref{table:rings}). The best fit model also includes a small ring of 4 au in radius, therefore suggesting the  presence of a inner circular cavity like those observed in transitional disks.  However, as discussed in the next section, our azimuthally symmetric best fit model poorly matches the dust emission from the innermost disk regions which instead seems to indicate the presence of a dust crescent.
 
 Finally, the last component of the model is a ring with a radius of about 32 au that is required to reproduce the kink in the radial profile of the emission observed at about 30 au from the center. Similar kinks in the radial profiled of the dust continuum emission are observed in other DSHARP sources and are discussed in Huang et al. (2018).

\subsection{Asymmetries of the continuum emission} 
\label{sec:contasym}
\begin{figure*}[t]
\includegraphics[width=\linewidth]{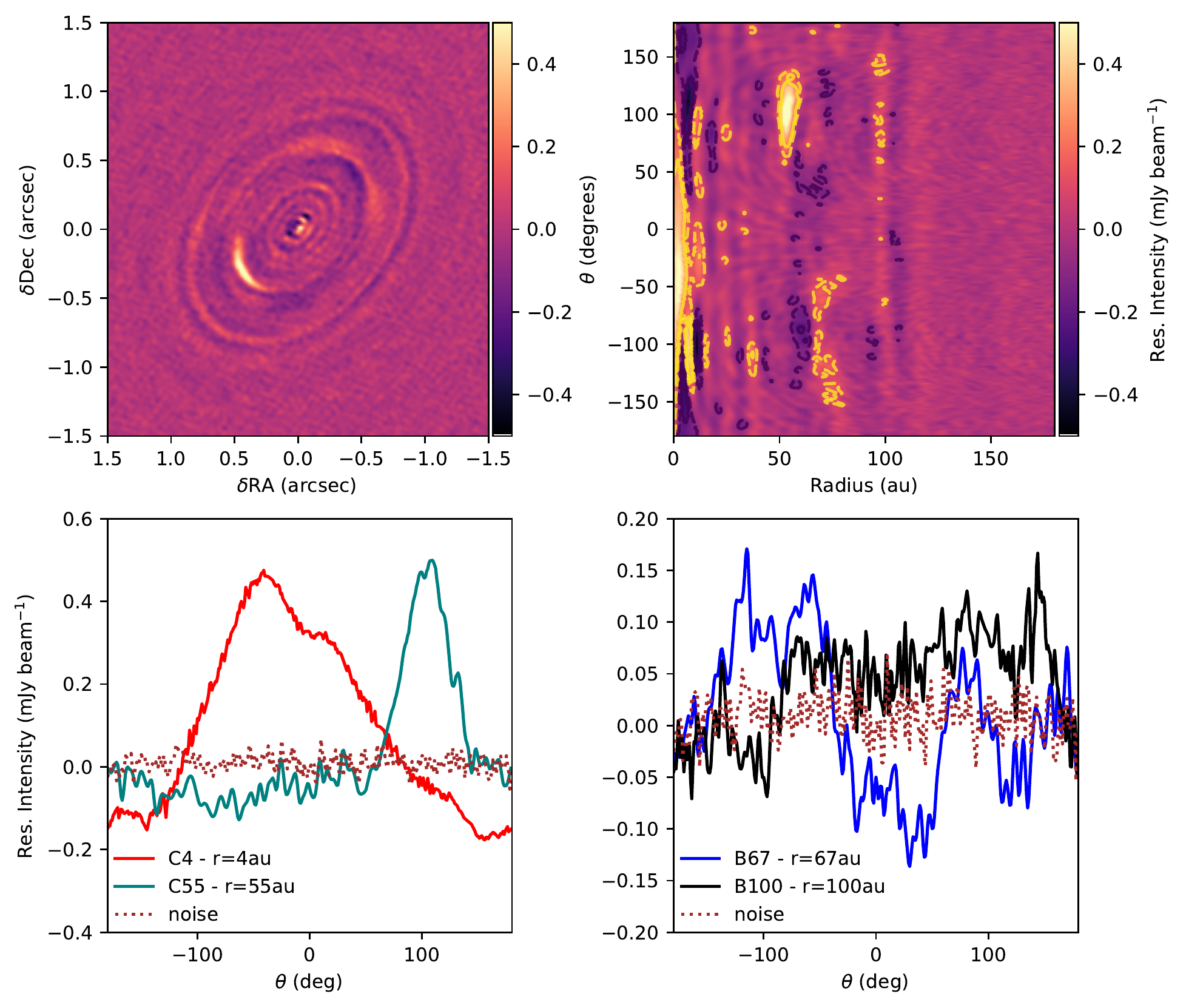}
\caption{\label{fig:fitvis} Top left and top right: Cartesian and polar maps of the residual intensity obtained by modeling the continuum image of HD~163296 with  a symmetric disk model comprising Gaussian rings as discussed in the text. Dashed and solid contours correspond to $\pm5\times$ and $\pm10\times$ the noise level, respectively. The bottom panels show the azimuthal profile of the residual intensity along the crescents C4 and C55 (left) and the rings B67 and B100 (right) . As a reference, the brown dotted curves indicate the azimuthal profile of the intensity at a radius of 160 au, which is indicative of the noise level of the observations.}
\end{figure*}

The map obtained by imaging the residual visibilities obtained by subtracting the best fit Gaussian model from the observations (Figure~\ref{fig:fitvis}) reveals several features with absolute intensities above 5 times the rms noise and peak intensities as high as 30 times the rms noise. The two most prominent structures have the shape of crescents and were already introduced at the beginning of Section~\ref{sec:cont} (see panels b and c of Figure 1). The innermost  crescent is located at about 4 au from the center and it overlaps with the innermost ring component of the Gaussian model discussed above. Given its azimuthally asymmetric structure, and following the nomenclature adopted for the rings and gaps, we label this feature as C4 (C meaning ``crescent''). The peak of C4 has a position angle of -44\arcdeg\, and an intensity is 0.55 mJy beam$^{-1}$, or about 20\% of the continuum intensity measured at the same position. C4 extends by about 180\arcdeg\, in azimuth and by about one resolution element in radius. C4 resembles the crescents observed in other Herbig Ae disks  that have been attributed to the presence of anticyclonic vortices \citep[e.g.][]{casassus13, isella13, vandermarel13,  boehler18} or to an optically thick inner disk warped with respect to the outer disk \citep[e.g.][]{marino15,benisty17}.  However, higher angular resolution observations are required to investigate the morphology of HD~163296 innermost disk regions in greater details.

The second crescent, labelled as C55, is centered at an orbital radius of about 55 au and position angle of 99\arcdeg. The peak intensity is 0.64 mJy beam$^{-1}$, corresponding to a signal-to-noise ratio of 28.
Modeling C55 intensity as a 2D Gaussian function in the polar plane $I(r, \theta) \propto e^{-[(r-r_c)^2/2\sigma_r^2+(\theta-\theta_c)^2/2\sigma_\theta^2]}$, we find $\sigma_\theta = 17$ au and $\sigma_r = 3$ au. Accounting for beam convolution (Equation 3), the intrinsic radial and azimuthal widths of C55 are $\sim2.2$~au and $\sim16.9$~au, respectively. 

In addition to C4 and C55, the image of the residuals shows fainter asymmetries along the B67 and B100 dust rings. If the dust emission is optically thin, intensity variations along the rings might probe the clumpiness of the dust distribution, while if the emission is optically thick, they inform about variations in the dust temperature. The intensity profiles along B67 and B100 are shown in the bottom right panel of Figure~\ref{fig:fitvis}. In the case of B67, the residual intensity varies between $\pm$0.15 mJy beam$^{-1}$,  to be compared with an azimuthally averaged intensity value measured at the same radius of 1.0 mJy beam$^{-1}$. The azimuthal intensity  profile of B67 shows small amplitude variations characterized by an angular scale of about 4\arcdeg, and larger variations with angular scales of 30-50\arcdeg. Whereas the small scale variations have the same size of the synthesized beam and are most likely caused by the noise of the observations, the larger scale structures might trace real asymmetries in the dust emission and, since the emission from B67 is estimated to be optically thin (see Section~\ref{sec:scat}), in the dust density.

\section{$^{12}$CO J=2-1 line emission}
\label{sec:co}

\begin{figure*}[t]
\centering
\includegraphics[width=0.9\textwidth]{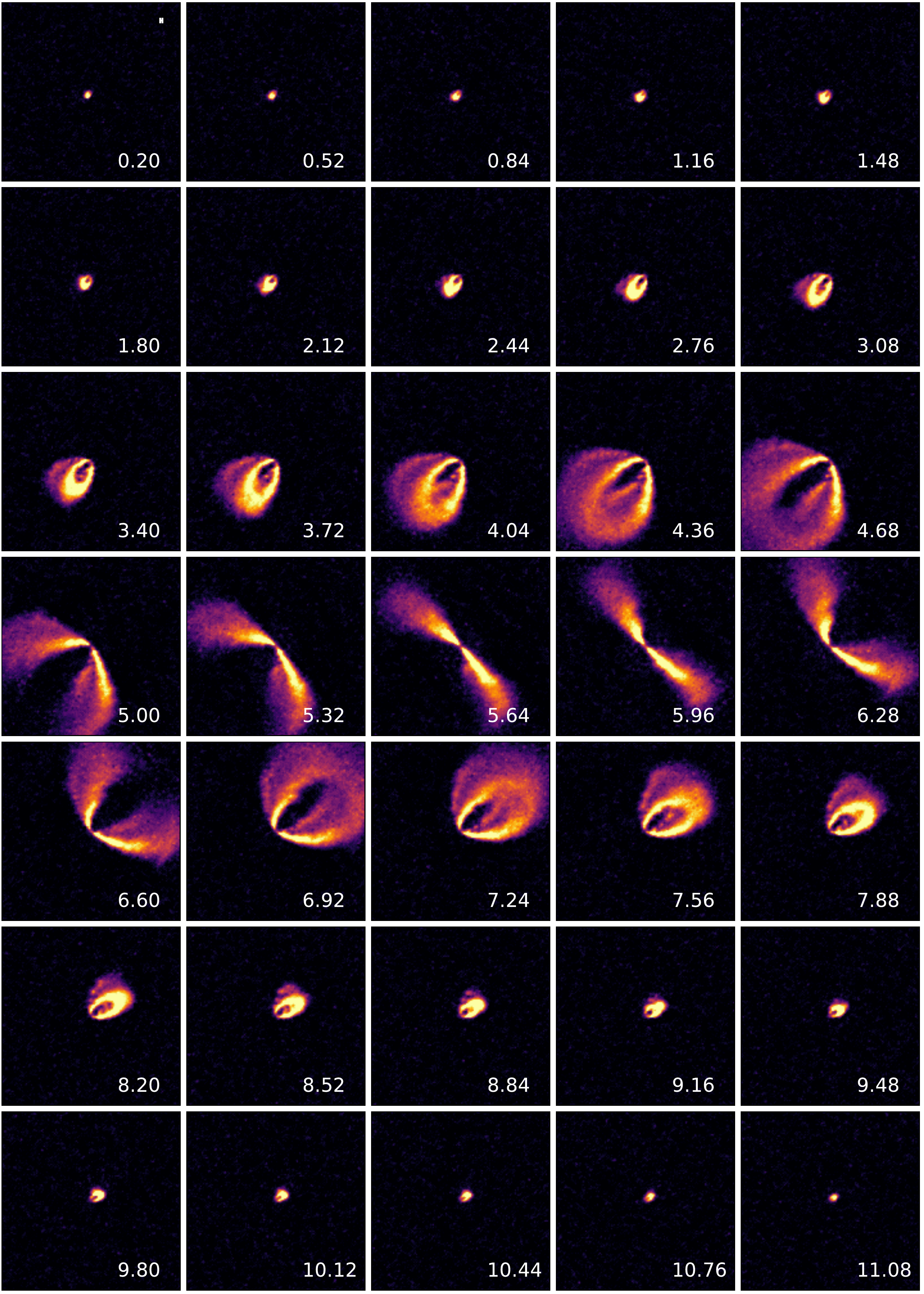}
\caption{Map of the $^{12}$CO J=2-1 line emission recorded toward HD~163296. The FWHM of the synthesized beam is 0.104\arcsec$\times$0.095\arcsec, corresponding to a spatial scale of about 10 au at the distance of the source. Each panel has a size of 8\arcsec$\times$8\arcsec, and is labeled with the velocity relative to the local standard of rest ($v_{lsr}$). Channels are spaced by 0.32 km s$^{-1}$. The color scale of the map is linear. \label{fig:comaps}}
\end{figure*}

Figure~\ref{fig:comaps} shows continuum-subtracted channels maps of the $^{12}$CO (2-1) emission (hereafter referred simply as CO emission). 
Below we describe the morphology and kinematics of the CO emission, and use the observations to constrain disk properties such as the disk temperature and the optical depth of the dust continuum emission. 

\subsection{Morphology of the CO emission}
\label{sec:co_morph}

The CO emission extends far beyond the outer edge of millimeter-wave continuum emission up to an angular distance of 5.6\arcsec\, or about 560 au, from the central star. As discussed in \cite{isella16}, the discrepancy between the radial extent of the millimeter wave dust and CO emission indicates a sharp drop in either the dust opacity or dust column density beyond about 200 au. 

\begin{figure*}[t]
\centering
\includegraphics[width=0.8\textwidth]{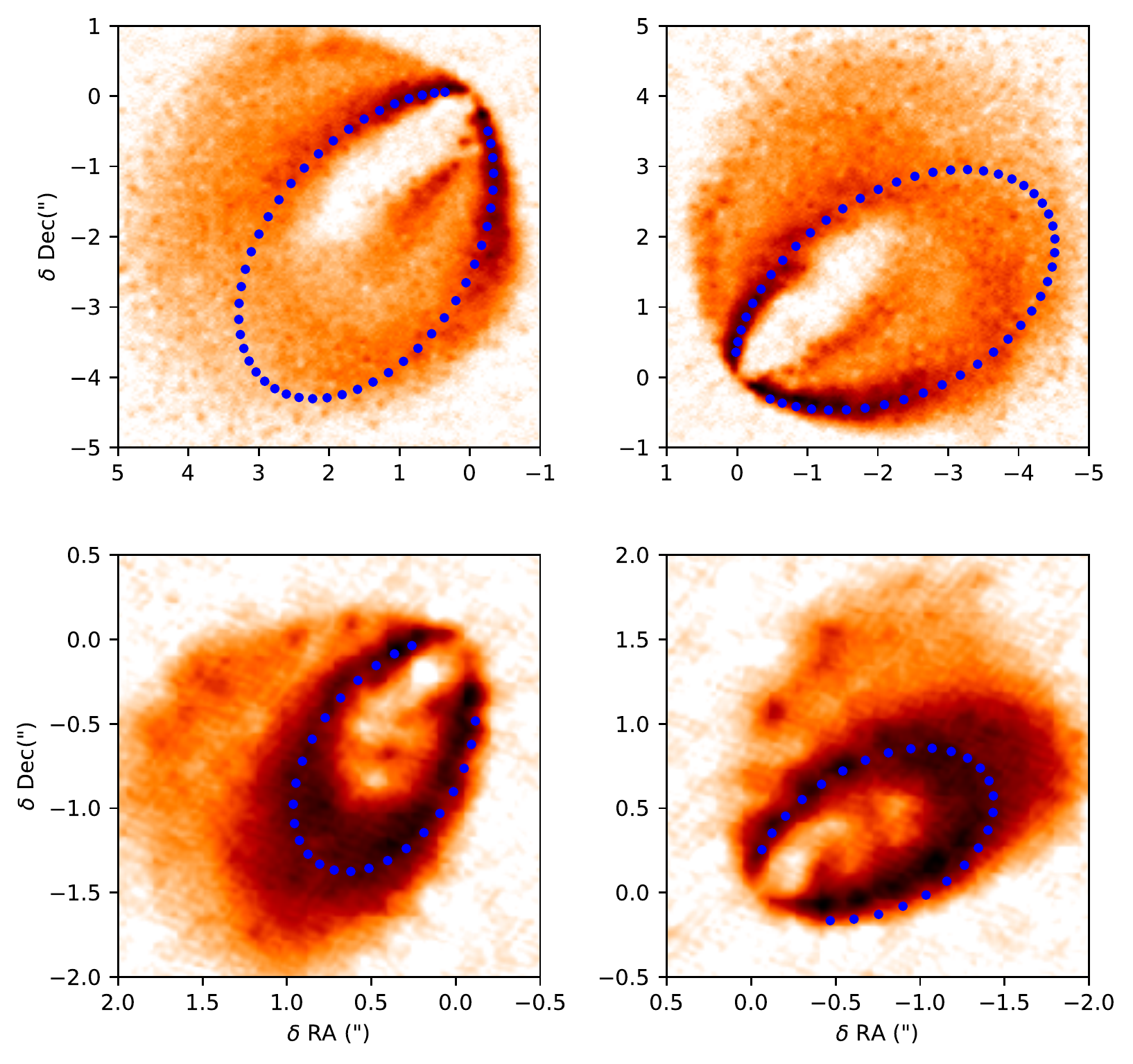}
\caption{Zoom-in of four of the CO channels shown in the previous figure (top left: $v_{lsr}=4.68$ km s$^{-1}$, top right:$v_{lsr}=6.92$ km s$^{-1}$, bottom left: $v_{lsr}=3.4$ km s$^{-1}$, bottom right:$v_{lsr}=8.2$ km s$^{-1}$ ). Blue dots indicate the isovelocity curves relative to the front CO emitting layer \label{fig:co_surf}}
\end{figure*}

The CO emission observed at intermediate velocities  (e.g., $v_{lsr} = 3.40-5.00$ km s$^{-1}$ and $v_{lsr} = 6.28-8.20$ km s$^{-1}$) shows a characteristic ``dragonfly" structure tracing both the front and rear warm CO emitting layers of the disk (Figure~\ref{fig:co_surf}). The emission coming from the rear of the disk is fainter than that from the front due to a combination of lower temperature of the emitting gas and absorption from intervening material. As discussed in \cite{rosenfeld13a}, the fact that the front and rear sides of the disk appear as two distinct emitting regions is explained by the large optical depth of the CO line (whose intensity therefore mostly depend on the gas temperature) and a vertical gradient of the disk temperature characterized by warm surfaces and a cold disk midplane. Furthermore, the separation between the front and rear CO emission measures the vertical geometry of the CO emitting surfaces. 

We compare the observed geometry of the CO emission with that of a parametric model where the CO line originates from a geometrically thin layers characterized by a distance from the midplane $z_{co} = \pm z_{co,0} (r/r_0)^q$. As demonstration of the technique, we show in Figure~\ref{fig:co_surf} parametric isovelocity curves corresponding to the front side of the disk calculated as presented in the Appendix A. The geometry of the isovelocity curves depends on $z_{co,0}$, $q$, as well as on the stellar mass and disk inclination. Assuming the disk inclination derived from the continuum ($i=46.7$\arcdeg) and a stellar mass of 2.0~M$_\star$, we obtain a good match between model and observations for $q=0.5$ and $z_{co}(100\mathrm{au}) = 30$~au.   

\begin{figure*}[t]
\centering
\includegraphics[width=1.0\textwidth]{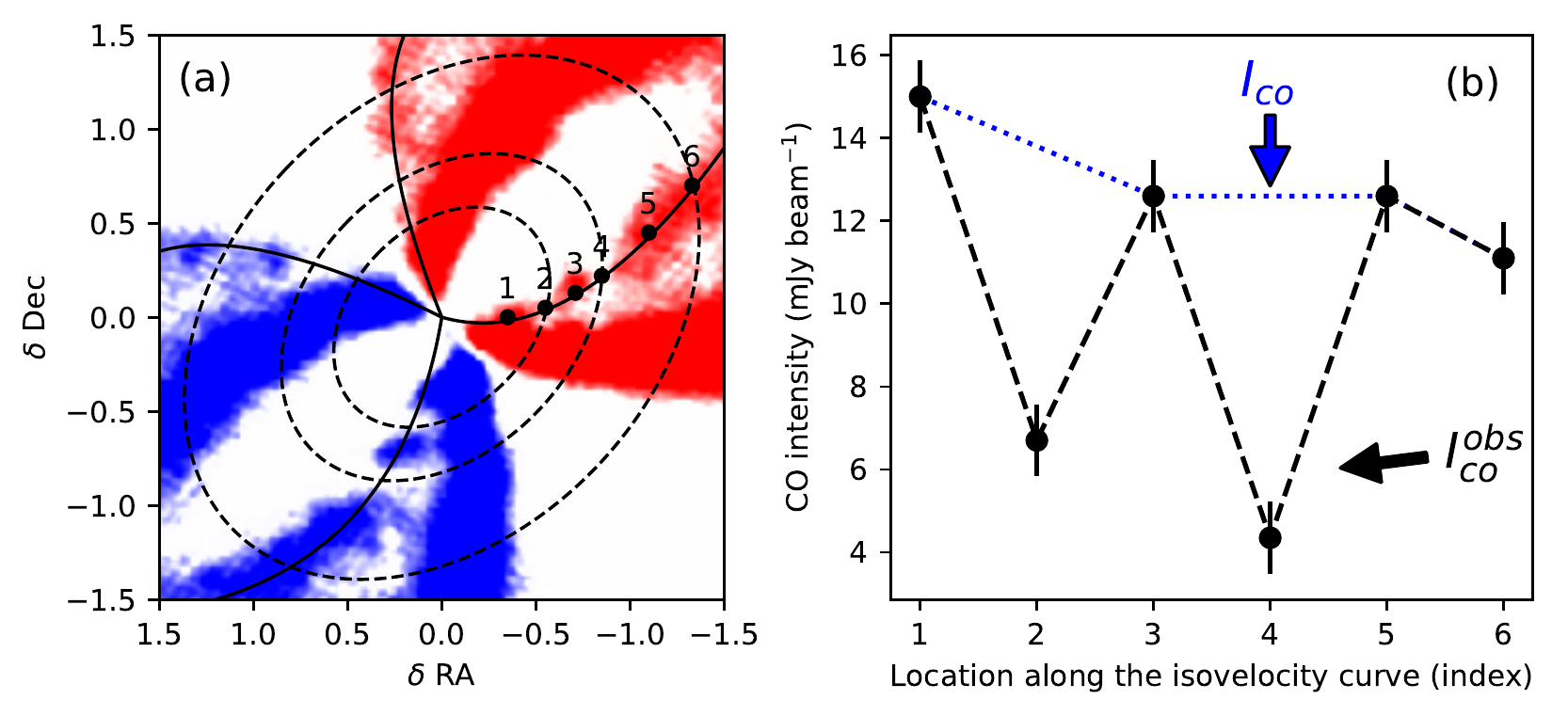}
\caption{Illustration of the technique used to measure the extinction optical depth of the dusty rings from observations of the CO emission. Panel (a): blue and red colors correspond to CO emission recorded at $v_{lsr}=4.36$ and $v_{lsr}=7.24$ km s$^{-1}$, respectively. Dashed ellipses indicate the position of the rings B67, B100, and B160. Solid lines indicate the iso-velocity curves corresponding to the emission coming from the back side of the disk. Panel (b): filled circles with error bars indicate the intensity of the CO line measured along the isovelocity curve. The indices 1-6 on the x-axis correspond to the positions 1-6 along the curve shown in the right panel. The blue dashed line indicates the unattenuated CO emission.  \label{fig:co_tau}}
\end{figure*}

Our observations reveal that the CO emission coming from the back side of the disk is dimmer at the location of the bright dust rings B67 and B100 compared to the emission observed at the location of the dark gaps D45 and D87 (Figure~\ref{fig:co_tau}). This dimming, which is likely due to absorption of the line emission as it crosses the dusty rings, provides a tool to constrain the optical depth of continuum emission independently from the assumptions on the dust temperature. The CO emission coming from the back side of the disk can be expressed as $I_{co}^{obs} = I_{co}e^{-\tau_d^{ext}}$, where $\tau_d^{ext}$ is the line of sight dust extinction optical depth equal to the sum of the absorption and scattering optical depths ($\tau^{ext}_d = \tau^{abs}_d+\tau^{sca}_d$), while $I_{co}$ is the initial unattenuated CO emission.  
Defining the parameter $\epsilon$ as $\epsilon \equiv \tau_d^{abs}/\tau_d^{ext}$, the absorption optical depth of the dust continuum  can be 
written as

\begin{equation}
\label{eq:tau_co}
\tau_d^{abs} = \epsilon  \ln{\frac{I_{co}}{I_{co}^{obs}}}
\end{equation}

The observations of the CO emission directly measure $I_{co}^{obs}$ along the bright continuum rings, while the unattenuated emission $I_{co}$ can be estimated through the procedure illustrated in Figure~\ref{fig:co_tau}. In practice, in the channels where CO absorption is observed, we measure the intensity of the CO emission along the isovelocity curves corresponding to the back side emission. We then use the values measured within the dust gaps, where, based on the results of the previous section, $\tau_d \ll 1$, to estimate $I_{co}$ at the position of the bright rings. We measure the ratio $I_{co}/I_{co}^{obs}$ for B67 and B100 along 18 different lines of sight (i.e., 18 different azimuthal angles along the dust rings) and find mean values of 1.9$\pm$0.1 and 2.1$\pm$0.1. We also perform this measurement for B168 finding $I_{co}/I_{co}^{obs}=1.1\pm0.2$. The corresponding dust extinction optical depths are $\tau^{ext}_d(\mathrm{B}67)=0.64\pm0.05$, $\tau^{ext}_d(\mathrm{B}100)=0.74\pm0.05$, $\tau^{ext}_d(\mathrm{B}168)=0.1\pm0.2$. It is worth noting that the ratio $I_{co}/I_{co}^{obs}$ could vary with the azimuthal angle if $\tau_d^{ext}$ is not constant along the dust rings. However, the sensitivity of our observations does not allow to perform such an analysis, and we must instead rely on the azimuthally averaged value of the CO intensity ratio to investigate the dust extinction properties.

As from Equation~\ref{eq:tau_co}, the absorpion optical depth of the dust rings depends on the dust scattering properties through the parameter $\epsilon$. For dust grains much smaller than the wavelength of the observations (in our case, for particle sizes $a \ll \lambda/2\pi \sim 0.2$~mm), scattering is negligible compared to absorption, and $\epsilon \simeq 1$. For grain sizes comparable to the wavelength of the observations ($a \sim \lambda/2\pi \sim 0.2$~mm), scattering and absorption properties of spherical grains follow Mie theory, and $\epsilon$ can vary between about 0.1 and 1, depending on the dust composition and internal structure. 
Finally, if $a \gg 0.2$~mm, scattering and absorption are equal, and  $\epsilon=0.5$ \citep[see, e.g.,][]{bohren98}. Without any prior information about the size and composition of dust grains, the observed absorption of the CO emission constrains the absorption optical depth of B67 and B100 between about 0.07 and 0.7, while the lack of CO absorption at the location of B168 sets an upper limit to the absorption optical depth of this ring at about 0.3. A caveat of this approach is that the emission from the front side of the disk could partially contaminate the emission from the back side, particularly close to the star where the separation between the two surfaces is small. Also, the ratio $I_{co}/I_{co}^{obs}$ could be affected by beam smearing, implying that our measurements provide only a lower limit for the true contrast between the initial and absorbed CO emission, and consequently, the corresponding dust optical depth. However, the angular resolution of the CO observations is sufficiently high to believe that this effect is small.  

If the absorption optical depth of the dust rings is measured from CO absorption, the corresponding continuum intensity $I_\nu$ can be used to estimate the temperature $T_d$ of the emitting dust.  If dust scattering is negligible ($\tau_d^{ext}=\tau_d^{abs}$ and $\epsilon=1$), the conversion between flux density and dust temperature is 
\begin{equation}
\label{eq:tdust}
    T_d = \frac{h \nu}{k_b} \left[ \ln \left( \frac{2 h \nu^3 \Omega (1-e^{-\tau_d^{abs}})}{c^2 F_\nu}  +1 \right) \right]^{-1},
\end{equation}
or, using Equation~\ref{eq:tau_co}, as 
\begin{equation}
\label{eq:tdust}
    T_d = \frac{h \nu}{k_b} \left[ \ln \left( \frac{2 h \nu^3 \Omega (1-I_{co}^{obs}/I_{co})}{c^2 F_\nu}  +1 \right) \right]^{-1}.
\end{equation}

From the measured continuum intensities we obtain dust temperatures of 24~K and 15~K for B67 and B100, respectively, while the upper limits for the optical depth of B168 translates in a minimum temperature of about 6 K. 

If dust scattering is not negligible ($\tau_d^{ext} > \tau_d^{abs}$ and  $\epsilon < 1$), the dust temperature can be estimated using the formalism presented in Birnstiel et al. (2018). In this case, the intensity is  written as $I_{\nu,d}\sim S(T_d, \tau_{\nu, d}^{ext}, \epsilon)(1-e^{-\tau_{\nu, d}^{ext}})$, where the source function $S_\nu$ depends on the dust temperature, and on the extinction and scattering optical depth along the line of sight. Adopting $\epsilon$ in the range between  0.1 and 1, we obtain dust temperatures between 130~K and 24~K for B67, and between 70~K and 15~K for B100. The fact that the dust temperature depends on the dust scattering properties hampers the application of this technique as an independent measure of disk temperature. However, if the dust temperature is measured through other means, then the dust scattering properties can by constrained (see Section~\ref{sec:discussion}).

\subsection{Gas temperature}
\label{sec:gastemp}

\begin{figure*}[t]
\centering
\includegraphics[width=1.0\textwidth]{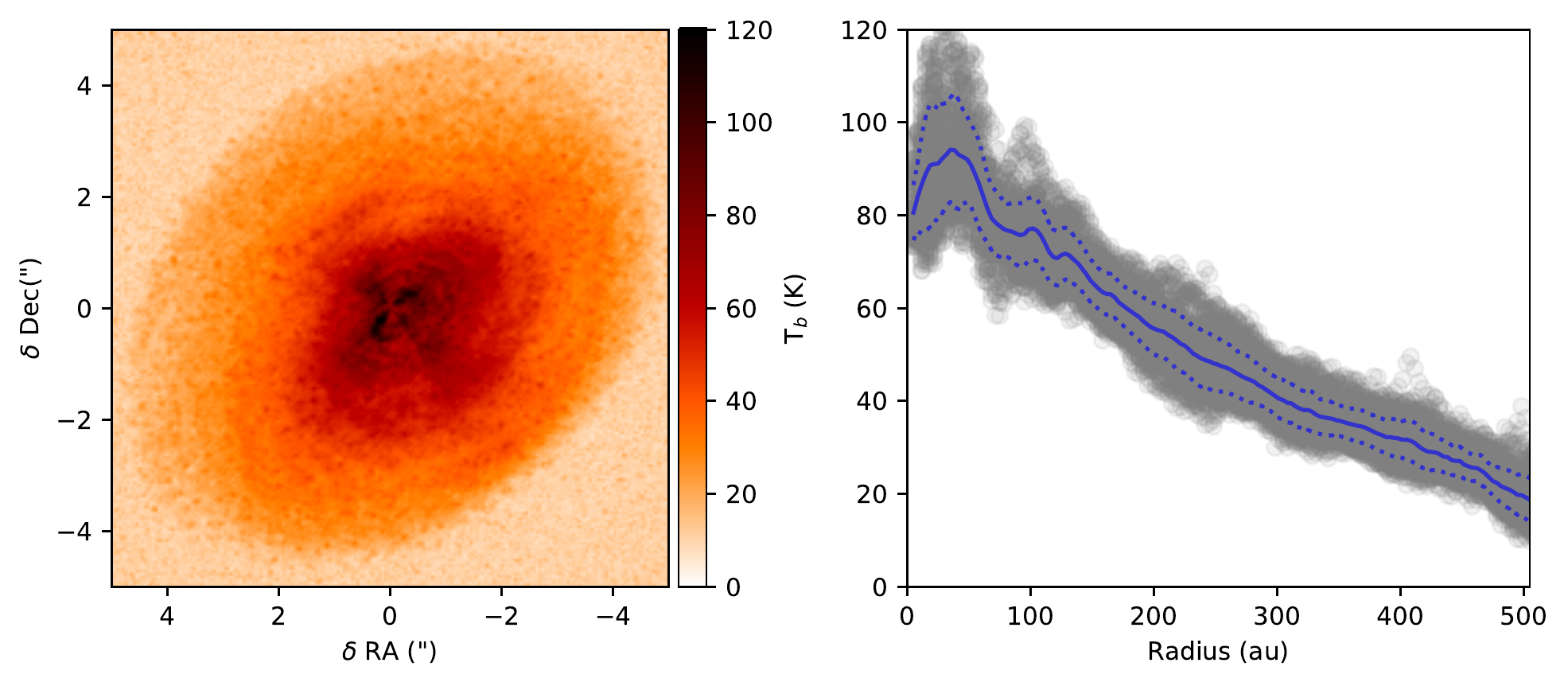}
\caption{Left: Peak intensity map of the $^{12}$CO J=2-1 line emission. Right: radial profile of the CO temperature deprojected accounting for the geometry of the emitting region as discussed in the text. Gray points show the scatter of the temperature measurements. The blue solid curve corresponds to the mean, while the dotted curve indicate the standard deviation from the mean. \label{fig:cotemp}}
\end{figure*}

At typical densities and temperatures of protoplanetary disks, the peak of the $^{12}$CO J=2-1 line emission is optically thick and the CO rotational levels are in local thermodynamic equilibrium \citep[see, e.g., ][]{weaver18}. The intensity at the peak of the line therefore measures the temperature of the emitting gas. Following \cite{weaver18}, we generate a CO temperature map from  non-continuum subtracted line emission to avoid underestimating the line intensity along line of sights where the continuum is optically thick (Figure~\ref{fig:cotemp}). The CO temperature reaches values as high as 120 K in the innermost disk regions and as low as 20 K at the disk outer edge. The peak intensity map is characterized by an evident asymmetry with respect to the apparent disk major axis: the south-west side of the disk is hotter than the north-east side. This is explained in first approximation by the fact that the CO emitting layer is a conic surface observed from a direction inclined with respect to the axis of the cone. The orientation of the asymmetry implies that the northern side of the disk is the closest to the observer \citep[see, also,][]{rosenfeld13a}. 

Assuming, as in the previous section, that the peak of the line is emitted from a geometrically thin layer at a distance from the disk midplane given by $z_{co}=z_{{co},0}(r/r_0)^q$, it is possible to deproject the observed CO temperature and estimate its profile as a function of the cylindrical radius $r$. Using the values for $z_{co}$ and $q$ presented above, we derive the CO temperature profile shown in the right panel of Figure~\ref{fig:cotemp}. Between 30 and 500 au, the radial profile of the CO temperature scales almost linearly with the radius and follows the relation $T_{co} (\mathrm{K}) \sim 87 - 0.14 (r/\mathrm{au})$. Within 30 au, the CO brightness temperature drops. This is consistent with the expected drop in brightness temperature as the emitting area becomes smaller than the angular resolution of the observations (beam dilution).

\subsection{Kinematics of the CO emission}

\begin{figure*}[t]
\centering
\includegraphics[width=1.0\textwidth]{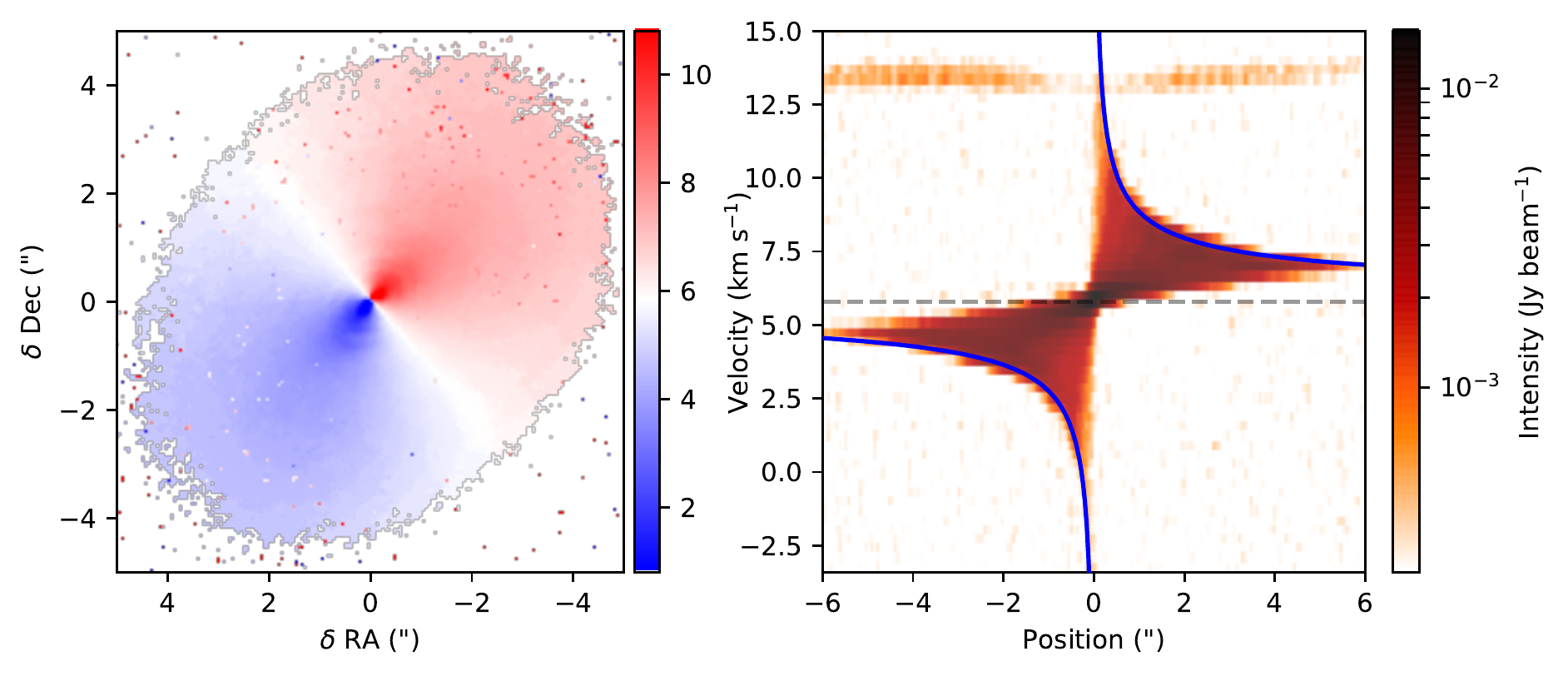}
\caption{Left: Intensity weighted velocity (moment I) of the $^{12}$CO J=2-1 line emission calculated by adopting an intensity threshold of 3 times the noise level. Right: Position-Velocity diagram obtained by averaging along the whole apparent disk minor axis. The position is the angular distance from the star along the projected disk major axis. The gray dashed line indicate the systemic velocity of 5.8 km s$^{-1}$. The solid blue curve corresponds to the maximum velocity along the line of sight for a Keplerian disk inclined by 46.7\arcdeg\, orbiting a 2.0~M$_\odot$ star at the distance of 101 pc. \label{fig:mom1_vp}}
\end{figure*}

The overall kinematics of the CO gas is well described by Keplerian rotation around a star with a mass of about 2.0~M$_\odot$ (Figure~\ref{fig:mom1_vp}). Both the integrated velocity map and the position-velocity diagram are in first approximation symmetric with respect to the apparent disk minor axis. The systemic velocity inferred from CO observations is 5.8 km s$^{-1}$ and is in agreement with previous results. The velocity-position diagram also reveal diffuse emission at a velocities between about 12.5 and 14 km s$^{-1}$ that might trace a redshifted stellar wind \cite[][]{klaassen13} or perhaps emission from the left-over of the HD~163296 parent cloud.

The CO map registered at a velocity of 6.92 km s$^{-1}$ (Figure~\ref{fig:comaps} and Figure~\ref{fig:co_surf}) is characterized by a kink at $\delta$RA$\sim$-1\arcsec,$\delta$Dec$\sim1.5$\arcsec. This feature was previously reported by \cite{pinte18} and has been attributed to perturbations in the gas kinematics caused by a planet with a mass of 2 M$_J$ orbiting at 260 au from the star.  Unfortunately, despite the better angular resolution, the observations obtained using the extended ALMA configuration achieve a spectral resolution about 10 times worse than that of those presented by \cite{pinte18}, and do not allow us to investigate the origin of this feature. Similarly, the coarse velocity resolution of our data might not allow to investigate the small deviations from keplerian rotation observed within the dust gaps and reported by \cite{teague18}.


\section{Discussion} \label{sec:discussion}

The two main results obtained so far are that (i) the combination of continuum and CO observations directly inform about the temperature and geometry of the HD~163296 disk and (ii) the analysis of the continuum emission constrains radial profile and level of asymmetry of the observed dust rings. In this section, we firstly compare the disk temperature inferred from the observations with current models for HD~163296 and, secondly, we discuss the possible origin of the morphology of the gas and dust emission. 

\subsection{Disk temperature and dust scattering}
\label{sec:scat}
\begin{figure*}[t]
\centering
\includegraphics[width=1.0\textwidth]{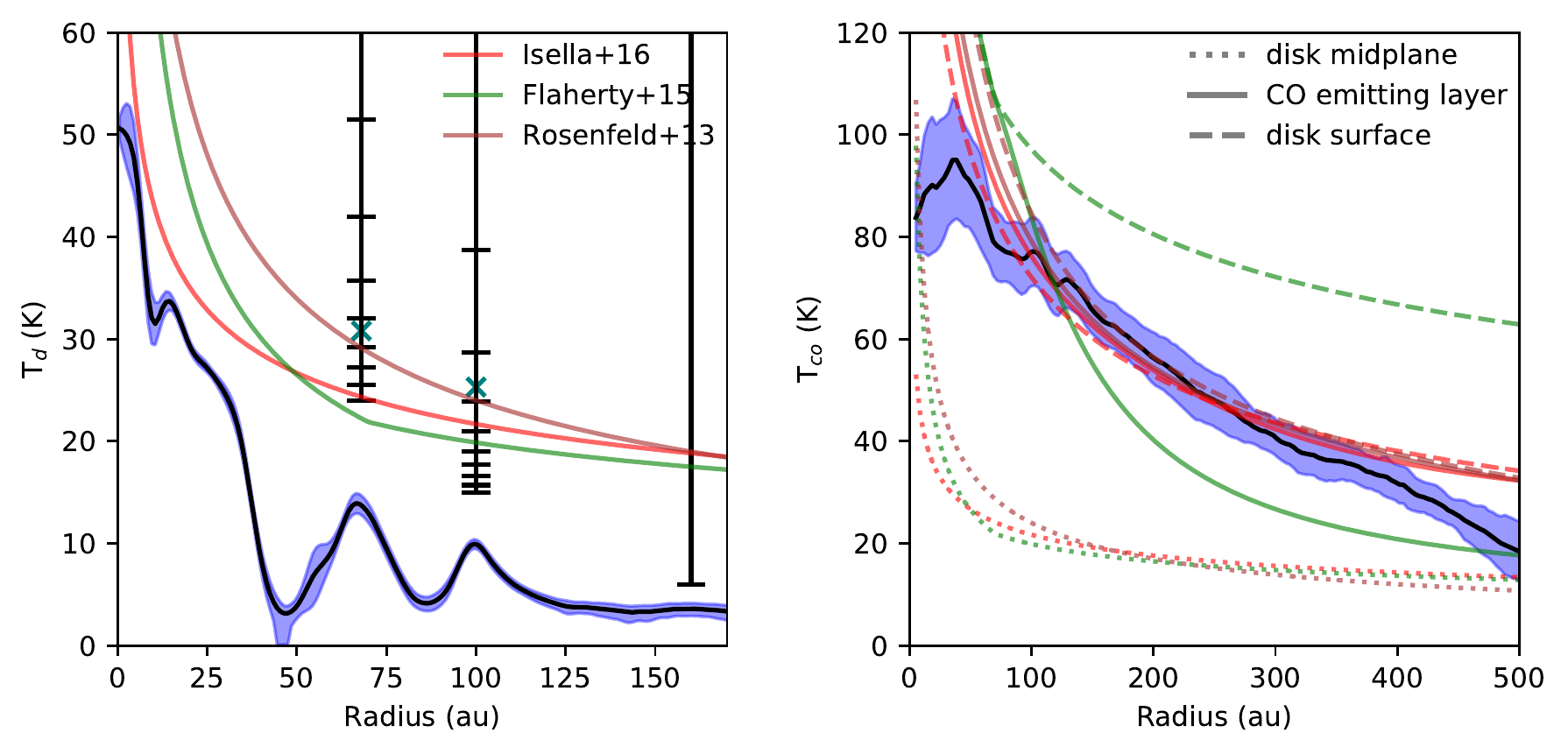}
\caption{\label{fig:comp_temp}Left: Dust temperature as a function of the orbital radius. The black and blue dashed region indicate the continuum brightness temperature and the standard deviation from the mean, respectively. Green, red and brown curves indicate the disk midplane temperature predicted by  \cite{flaherty15}, \cite{rosenfeld13a}, and \cite{isella16}, respectively. The vertical lines show the range of dust temperatures compatible with the observed absorption of the CO emission arising from the back side of the disk at the location of the bright rings B67, B100, and B168. The horizontal segments indicate temperatures for different assumptions on the dust scattering. From the bottom up, they correspond to $\epsilon=1$ (no scattering), 0.9, 0.8, 0.7, 0.6, 0.5, 0.4, and 0.3. For completeness, we also show with teal crosses the dust temperature assumed by Dullemond et al. (2018).  Right: The black curve and the dashed blue region indicate the CO peak brightness temperature and the standard deviation from the mean, respectively. Green, red and brown solid curves indicate the temperature of the CO emitting layer predicted by  \cite{flaherty15}, \cite{rosenfeld13a}, and \cite{isella16}. Dotted and dashed curves indicate the temperature in the disk midplane and disk surface predicted by the same models, as derived by modeling the dust continuum and $^{12}$CO J=2-1 emission.}
\end{figure*}

To date, the HD~163296 disk has been the subject of several studies aimed at constraining the density and thermal structure of the disk through forward modeling of the spatially resolved millimeter-wave molecular line and continuum emission \citep[e.g.,][]{isella07,isella16, rosenfeld13a, flaherty15, tilling12}. This approach relies on a large number of assumptions including the choice of the parametric forms for the gas and dust surface density and temperature (when the latter is not calculated using radiative transfer models), the dust and molecular abundances, the dust opacities, the stellar properties (mass, luminosity, distance), and  the gas kinematics (e.g. Keplerian motion, turbulent velocity). Because of the large computational time required to generate synthetic images at high spatial and spectral resolution, the comparison between models and observations is generally performed by either manually adjusting the model parameters until obtaining a satisfactory result \citep[e.g.][]{isella16}, or by using automatic algorithms to search for best fit models among a subset of model parameters. 
Not surprisingly, forward modeling performed by different investigators lead to different results. 

Figure~\ref{fig:comp_temp} compares the temperature inferred from 
the 1.25 mm dust continuum intensity to the models discussed in \cite{isella16}, \cite{rosenfeld13a}, and \cite{flaherty15}. We choose these three studies because they assume a similar parametric form for the disk temperature and are based on high spectral and angular resolution observations of the $^{12}$CO line emission. If scattering from dust grains is negligible ($\tau^{ext}_d = \tau^{abs}_d$, $\epsilon=1$), the brightness temperature  $T_b^{pl}$, the physical dust temperature $T_d$, and the optical depth $\tau^{abs}_d$ are linked by the ralation $B_\nu(T_b^{pl})=B_\nu(T_d)(1-e^{-\tau_d^{abs}})$. 
Within 30 au, the model temperatures are bracketed between \cite{rosenfeld13a} and \cite{isella16}, and would imply optical depths of about 1.3 and 0.7, respectively. Beyond 30 au, $T_b^{pl}$ drops well below the model predictions, suggesting that the emission becomes more optically thin. In particular, the model temperatures give  $\tau_d^{abs} < 0.02$ and $\tau_d^{abs} \sim 0.04-0.05$ at the center of the dark gaps D45 and D86, respectively, and $\tau_d^{abs} \sim 0.5-0.7$ and $\tau_d^{abs} \sim 0.3-0.5$ at the bright rings B67 and B100.

These measurements rely on the assumption that scattering is negligible, but, as discussed in Section~\ref{sec:co_morph}, this is likely not the case if dust grains have sizes comparable or larger than the wavelength of the observations. The inclusion of scattering significantly complicates the interpretation of the observed continuum emission and a simple relation between $T_b^{pl}$, $T_d$, and the absorption and scattering optical depths, $\tau^{abs}_d$ and $\tau^{sca}_d$ does not exist. Fortunately, however, it is possible to derive an approximated analytic solution to constrain $T_d$, $\tau^{abs}_d$ or $\tau^{sca}_d$ if the other two quantities are known (see Section 5 of Birnstiel et al. 2018). 

In the case of HD~163296, the direct measurement of the extinction opacity at the bright rings B67 and B100 discussed in Section~\ref{sec:co_morph} provide a unique tool to constrain the scattering properties of the circumstellar dust. Using the formalism discussed in Birnstiel et al. (2018), we calculate the physical dust temperature required to reproduce the dust continuum emission measured at B67 (0.77 mJy beam$^{-1}$) and B100 (0.45 mJy beam$^{-1}$) for values of the scattering parameter $\epsilon$ varying from 0.1 ($\tau_d^{sca}=9\tau_d^{abs}$) to 1 (no scattering).  The comparison between these temperatures, which are indicated with horizontal segments in the left panel of Figure~\ref{fig:comp_temp}, and those predicted by \cite{flaherty15}, \cite{isella16}, \cite{rosenfeld13a} constrain $\epsilon$, and therefore the dust scattering properties. 
For B67, we find that $\epsilon \gtrsim 0.6$ leads to dust temperature consistent with the models. We note that the midplane temperature of  \cite{flaherty15} is slightly below the minimum temperature consistent with the observations, which corresponds to $\epsilon=1$. Interestingly, for B100, the consistency between models and observations requires $\epsilon$ between 0.4 and 0.6, while $\epsilon=1$ would imply a physical dust temperature of only 15 K. The difference in scattering properties between B67 and B100 is intriguing and, taken at face value, might suggest a variation in the dust properties between B67 and B100. 
However, the difference might also result from the fact that, due to beam smearing, the measured value of $\tau_d^{ext}(\mathrm{B}67)$ might be a lower estimate of the extinction opacity, and consequently, of the scattering opacity. Future ALMA observations expressly designed to image the molecular gas emission from the HD~163296 at both high angular resolution and sensitivity are required to place more stringent constraints on the dust extinction optical depth. 

We conclude this section by comparing the peak brightness temperature of the CO emission measured in Section~\ref{sec:gastemp} to the temperature of the CO emitting layer predicted by \cite{flaherty15}, \cite{isella16}, and \cite{rosenfeld13a} models (see right panel of Figure~\ref{fig:comp_temp}). In Section~\ref{sec:co_morph} we found that the CO emitting layer corresponds to the surface defined by $z(r)=30 \mathrm{au} (r/100 \mathrm{au})^{0.5}$. Within about 100 au, the measured gas temperature is below the model predictions. This is likely due to the effect of beam dilution as discussed in \citet{weaver18}. Between 100 and 400 au, the temperature of the CO emitting layer predicted by \cite{rosenfeld13a} and \cite{isella16} is in good agreement with the observations, while the temperature predicted by \cite{flaherty15} is about 30\% lower. Also, while  the temperature of the CO emitting layer predicted by \cite{rosenfeld13a} and \cite{isella16} is similar to that of the stellar irradiated disk surface (dashed line), as expected in the case of a very optically thick line, \cite{flaherty15} temperature is closer to that achieved in the disk midplane (dotted lines). 



\subsection{On the origin of the dust rings}

\begin{figure*}[t]
\centering
\includegraphics[width=0.9\textwidth]{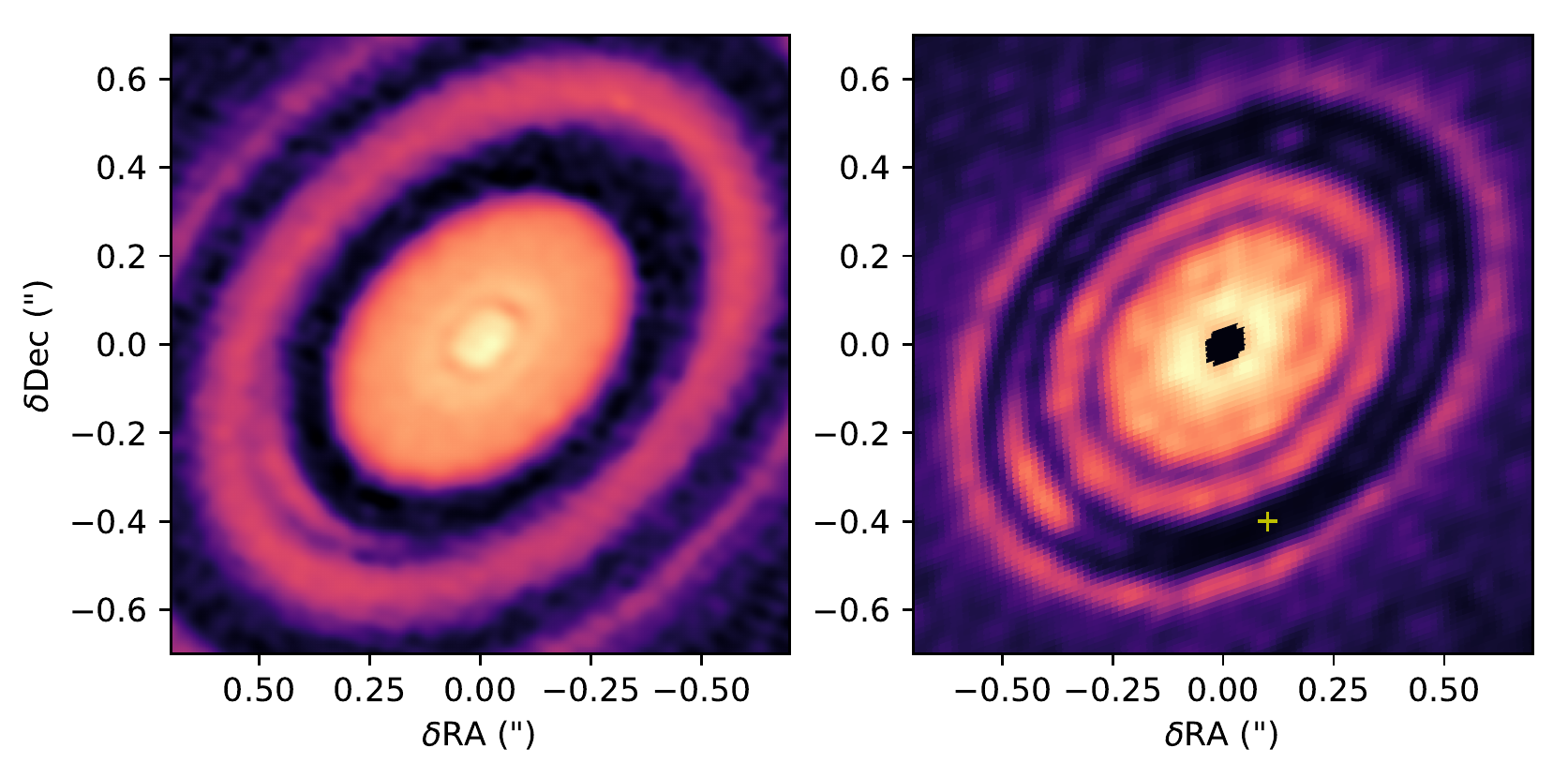}
\caption{\label{fig:zhang} Left: Image of the 1.25 mm dust continuum emission as in Figure 1. Right: Synthetic image of the 1.25 dust continuum emission of a disk perturbed by a 0.15 M$_J$ planet orbiting at about 54 au from the central star. The position of the planet is indicated by the yellow cross. The model is part of the suit of models presented in Zhang et al. (2018). The color scale of the two panels is the same and corresponds the intensity normalized to the peak value. No attempts have been made to match the model to the observations, beside rotating and inclining the model image to match the orientation of the HD~163296 disk.  }
\end{figure*}

In Section~\ref{sec:cont}, we found that the radial profile of the continuum emission from the HD~163296 disk can be represented by a sum of azimuthally symmetric Gaussian rings. However, significant asymmetric structures appear once the symmetric component of the emission is removed. In this section, we discuss these results in the framework of the planet-disk interaction mechanism, while we refer to Huang et al. (2018) for a discussion of other processes capable of generating dust rings. We also refer the reader to Dullemond et al. (2018) for a discussion of the observed rings properties in the context of dust trap models. 

Massive planets are expected to open gaps in the surface density of protoplanetary disks along their orbits through the strong mutual gravitational interaction with the circumstellar material \cite[see, e.g.,][]{bryden99}. While the planet minimum mass for this to happen varies as a function of the disk physical properties, most of the simulations agree that for typical disk conditions planets more massive than 20-40 Earth masses should produce gaps observable at the spatial scales probed by ALMA \citep[e.g.][]{dong12, isella18}. In the limit of nearly inviscid disks, simulations have also shown that even planets with masses as low as few Earth masses might generate visible rings \citep{dong18}. Given the ubiquity of planets in the Universe, it seems therefore reasonable to associate the ring structures observed by ALMA with the presence of planetary systems in the act of forming. 

In the case of HD~163296, the planet-disk interaction hypothesis is supported by the detection of gas depletion and deviation from Keplerian rotation within the dark gaps D87 and D143 \citep{isella16, teague18}, which imply that the emission gaps correspond to gaps in the gas density. The comparison with hydrodynamic simulations indicates that D45, D87 and D143 might results from the dust clearing operated by planets with masses between 0.5 and 2 M$_J$ \citep{liu18, teague18}. However, the ring morphology of the HD~163296 disk might also be consistent with gravitational perturbations by a single planet with a mass as low as 65 M$_E$ orbiting at about 100 au from the star if the disk viscosity is very low \citep[$\alpha<10^{-4}$,][]{dong18}.
Overall, these results are consistent with the planet masses discussed in Zhang et al. (2018), which have been estimated from the azimuthally averaged profile of the continuum emission shown in Figure~\ref{fig:cont_deproj}. In addition, Zhang et al. (2018) find that the newly discovered gap B10 might be generated by a planet with a mass between 0.2 and 1.5 M$_J$. 

Here we want to point out that the detection of asymmetries in the dust emission discussed in Section~\ref{sec:contasym} corroborates the hypothesis that the HD~163296 disk is shaped by the interaction with yet unseen planets.  As shown in Figure 6 and 7 of Zhang et al. (2018), features similar to the crescent observed inside the D45 gap (inset b of Figure~\ref{fig:cont})  are naturally produced by the disk-planet interaction whenever the disk viscosity is low ($\alpha < 10^{-3}$) and the planet mass is above about 0.1 M$_J$. The crescents observed in numerical simulations have two different origins: they either trace dust particles trapped in anticyclonic vortices that form at the edge of the gap opened by the planet \citep{li2001}, or probe material trapped at the Lagrangian points along the planet orbit. Without any pretense of accuracy, we compare in Figure~\ref{fig:zhang} the map of the 1.25~mm dust continuum  emission recorded toward HD~163296 with a synthetic image of a disk model characterized by the presence of a 0.15 M$_J$ planet orbiting at 54 au from the central star. The model is part of the suite of models presented in Zhang et al. (2018) and was generated assuming a viscosity parameter $\alpha=10^{-4}$ and a disk pressure scale height $h = 0.05 r$. This comparison is purely qualitative but nevertheless shows the resemblance between the corotation features predicted by models and the circular arc observed in HD~163296. 

\section{Conclusion} \label{sec:conc}

The unprecedented resolution of the DSHARP/ALMA data of the HD~163296 circumstellar disk enabled us to dive into 
a detailed characterization of the morphology of the dust continuum emission and  of some of the physical properties 
of the circumstellar material. The comparison between images of the 1.25~mm  continuum and $^{12}$CO line emission 
and parametric disk models led us to the following conclusions: 
\begin{itemize}
    \item The new ALMA observations confirm the presence of two bright rings in the millimeter-wave dust continuum emission centered at 67 and 101 au, and reveal an additional ring with a radius of 15 au. The radial profile of the dust emission across the rings is well described by a Gaussian profiles with widths of 8.7, 6.6, and 5.8 au, respectively. 
    \item The dust continuum emission is characterized by several asymmetric structures. The most prominent consist in crescents centered at about 4 and  55 au. We also find asymmetric structures along the ring B67 characterized by amplitude variations of $\pm15\%$  relative to the ring mean intensity. We argue that this asymmetries trace local variations in the dust densities and support the hypothesis that the HD~163296 disk is shaped by the gravitational interaction with yet unseen planets. 
    \item The observations of the $^{12}$CO line emission probe the temperature and geometrical structure of the emitting gas. We find that the 
    peak of the line arises from a disk layer at a vertical distance from the disk midplane given by  $z_{co}= 30\mathrm{au}(r/100\mathrm{au})^{0.5}$. Between 30 au and 500 au, the temperature of the CO emitting layer scales  linearly from the orbital radius and follows the relation $T_{co}(\mathrm{K})\sim87 - 0.14(r/\mathrm{au})$. 
    \item The CO maps show the line emission coming  from both the front and rear side of the disk. The latter is attenuated as it passes through the dust rings, providing a tool to directly measure the extinction optical depth $\tau_d^{ext}$ of the dust. We measure $\tau_d^{ext} \sim 0.7$ at the position of the B67 and B100 rings, and $\tau_d^{ext} < 0.3$ for B168. We point out that if the absorption optical depth 
    of the dust rings $\tau_d^{abs}$ can be measured through other means, a direct measurement of $\tau_d^{ext}$ allows to constrain the dust scattering properties.  
    \item Finally, we compare the CO temperature inferred from our ALMA observations to theoretical predictions based on previous forward 
    modeling of millimeter wave observations of HD~163296. Overall, we find that models predict temperatures within 30\% of the measured values. By adopting the predicted temperatures for the disk midplane, we find that $\tau_d^{abs}$ is between about 0.5 and 0.6 at B67, and between 0.3 and 0.5 at B100. This implies $\tau_d^{sca}$ between 0.1 and 0.2 at B67, and between 0.2 and 0.4 at B100.   
\end{itemize}

\acknowledgments  We thank the anonymous referee for the comments that helped improving the manuscript. A.I. acknowledges support from the National Aeronautics and Space Administration under grant No. NNX15AB06G issued through the Origins of Solar Systems program, and from the National Science Foundation under grant No.~AST-1715719. S.A. and J.H. acknowledge support from the National Aeronautics and Space Administration under grant No.~17-XRP17$\_$2-0012 issued through the Exoplanets Research Program.  J.H. acknowledges support from the National Science Foundation Graduate Research Fellowship under Grant No.~DGE-1144152.  L.P. acknowledges support from CONICYT project Basal AFB-170002 and from FCFM/U.~de Chile Fondo de Instalaci\'on Acad\'emica.   C.P.D. acknowledges support by the German Science Foundation (DFG) Research Unit FOR 2634, grants DU 414/22-1 and DU 414/23-1.  V.V.G. and J.C acknowledge support from the National Aeronautics and Space Administration under grant No.~15XRP15$\_$20140 issued through the Exoplanets Research Program.  T.B. acknowledges funding from the European Research Council (ERC) under the European Union’s Horizon 2020 research and innovation programme under grant agreement No.~714769. M.B. acknowledges funding from ANR of France under contract number ANR-16-CE31-0013 (Planet Forming disks).  Z.Z. and S.Z. acknowledge support from the National Aeronautics and Space Administration through the Astrophysics Theory Program with Grant No.~NNX17AK40G and the Sloan Research Fellowship.  L.R. acknowledges support from the ngVLA Community Studies program, coordinated by the National Radio Astronomy Observatory, which is a facility of the National Science Foundation operated under cooperative agreement by Associated Universities, Inc. M.B. acknowledges funding from ANR of France under contract number ANR-16-CE31-0013 (Planet Forming disks).

\section*{Appendix A: calculation of isovelocity lines}
The isovelocity lines shown in Figure~\ref{fig:co_surf} are calculated as follow. Firstly, we assume that CO molecules rotate around the star at the Keplerian velocity $v_k(R) = \sqrt{GM_\star/R}$, where $R$ is the spherical radius. Secondly, we assume that the line emission arises from a geometrically thin layer characterized by the distance from the disk midplane $z_{co}(r) = z_0(r/r_0)^q$, where $r$ is the cylindrical radius, $r=\sqrt{R^2+z_{co}^2}$. Finally, we assume that the disk midplane is inclined with respect to the line of sight by an angle $i$ and is rotated in the sky by the position angle PA. The component of the Keplerian velocity along the line of sight is $v_l(R,\theta) = v_k(R) \sin{i} \sin{\theta}$, where $\theta$ is the azimuthal angle. The isovelocity contours corresponding to the velocity $v$ in a spherical reference frame centered at the center of the disk is calculated by finding the radii that satisfy the relation $v - v_l(R,\theta) = 0$, for $\theta$ between 0 and 2$\pi$. 
Finally, the conversion between the reference frame centered on the disk and the image plane is 

\begin{eqnarray}
\delta \mathrm{RA} & = & x' \sin{\mathrm{PA}} - y' \cos{\mathrm{PA}} \\
\delta \mathrm{Dec} & = & x' \cos{\mathrm{PA}} + y' \sin{\mathrm{PA}}
\end{eqnarray}
where
\begin{eqnarray}
x' & = & r \cos{\theta} \cos{i} + z_{co}(r) \sin{i} \\
y' & = & r \sin\theta
\end{eqnarray}







\begin{figure*}[t]
\centering
\includegraphics[width=0.9\textwidth]{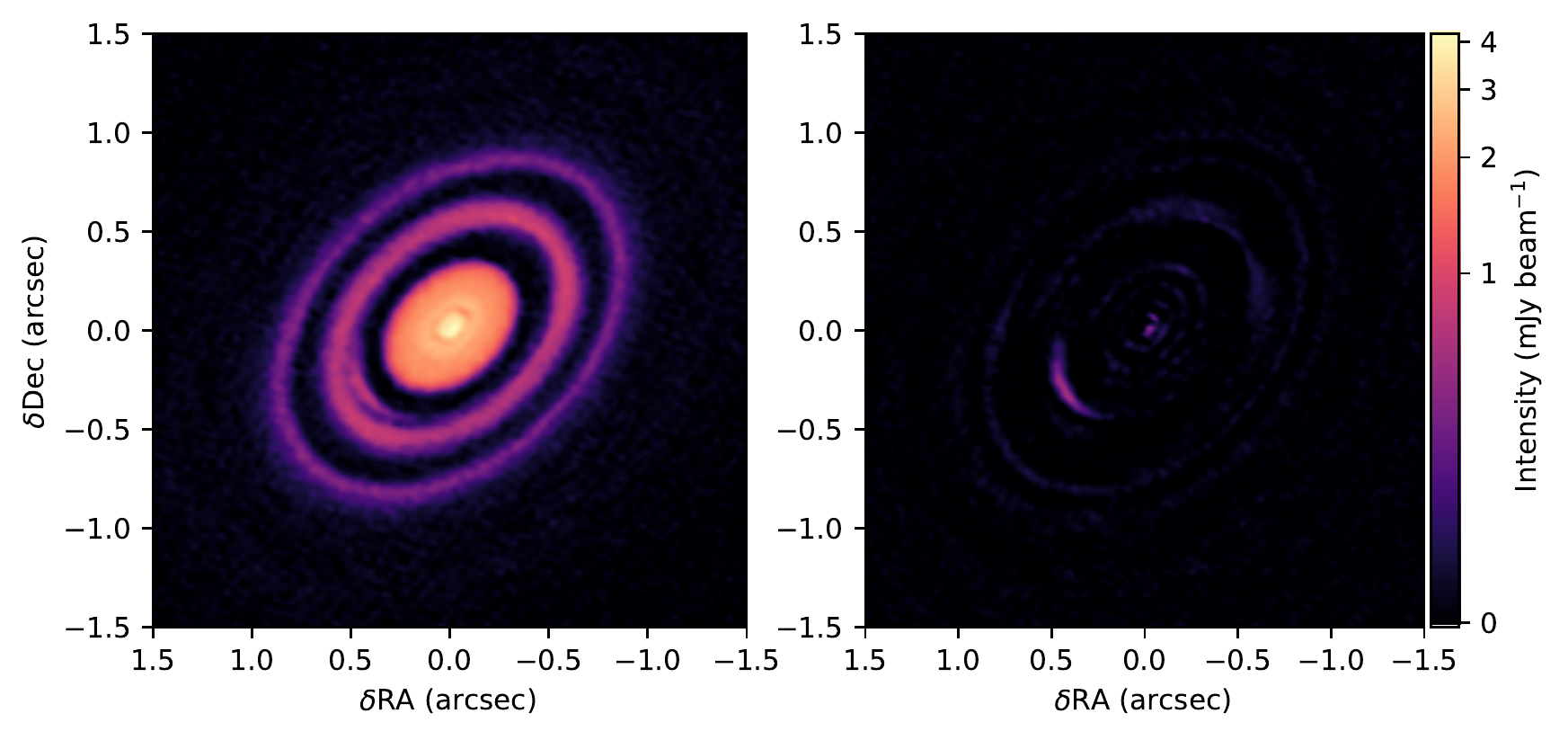}
\caption{\label{fig:res_comp} Left: Map of the 1.25 mm continuum emission as in Figure~\ref{fig:cont}. Right: Image of the residual visibilities resulting  from the Gaussian model fitting discussed in Section~\ref{cont:width}. Both panels use the same color scale to facilitate the comparison.}
\end{figure*}


\begin{thebibliography}{}
\expandafter\ifx\csname natexlab\endcsname\relax\def\natexlab#1{#1}\fi
\providecommand{\url}[1]{\href{#1}{#1}}
\providecommand{\dodoi}[1]{doi:~\href{http://doi.org/#1}{\nolinkurl{#1}}}
\providecommand{\doeprint}[1]{\href{http://ascl.net/#1}{\nolinkurl{http://ascl.net/#1}}}
\providecommand{\doarXiv}[1]{\href{https://arxiv.org/abs/#1}{\nolinkurl{https://arxiv.org/abs/#1}}}

\bibitem[{{ALMA Partnership} {et~al.}(2015){ALMA Partnership}, {Brogan},
  {P{\'e}rez}, {Hunter}, {Dent}, {Hales}, {Hills}, {Corder}, {Fomalont},
  {Vlahakis}, {Asaki}, {Barkats}, {Hirota}, {Hodge}, {Impellizzeri}, {Kneissl},
  {Liuzzo}, {Lucas}, {Marcelino}, {Matsushita}, {Nakanishi}, {Phillips},
  {Richards}, {Toledo}, {Aladro}, {Broguiere}, {Cortes}, {Cortes}, {Espada},
  {Galarza}, {Garcia-Appadoo}, {Guzman-Ramirez}, {Humphreys}, {Jung}, {Kameno},
  {Laing}, {Leon}, {Marconi}, {Mignano}, {Nikolic}, {Nyman}, {Radiszcz},
  {Remijan}, {Rod{\'o}n}, {Sawada}, {Takahashi}, {Tilanus}, {Vila Vilaro},
  {Watson}, {Wiklind}, {Akiyama}, {Chapillon}, {de Gregorio-Monsalvo}, {Di
  Francesco}, {Gueth}, {Kawamura}, {Lee}, {Nguyen Luong}, {Mangum}, {Pietu},
  {Sanhueza}, {Saigo}, {Takakuwa}, {Ubach}, {van Kempen}, {Wootten},
  {Castro-Carrizo}, {Francke}, {Gallardo}, {Garcia}, {Gonzalez}, {Hill},
  {Kaminski}, {Kurono}, {Liu}, {Lopez}, {Morales}, {Plarre}, {Schieven},
  {Testi}, {Videla}, {Villard}, {Andreani}, {Hibbard}, \&
  {Tatematsu}}]{brogan15}
{ALMA Partnership}, {Brogan}, C.~L., {P{\'e}rez}, L.~M., {et~al.} 2015, \apjl,
  808, L3, \dodoi{10.1088/2041-8205/808/1/L3}

\bibitem[{{Andrews} {et~al.}(2011{\natexlab{a}}){Andrews}, {Rosenfeld},
  {Wilner}, \& {Bremer}}]{andrews11b}
{Andrews}, S.~M., {Rosenfeld}, K.~A., {Wilner}, D.~J., \& {Bremer}, M.
  2011{\natexlab{a}}, \apjl, 742, L5, \dodoi{10.1088/2041-8205/742/1/L5}

\bibitem[{{Andrews} {et~al.}(2011{\natexlab{b}}){Andrews}, {Wilner},
  {Espaillat}, {Hughes}, {Dullemond}, {McClure}, {Qi}, \& {Brown}}]{andrews11}
{Andrews}, S.~M., {Wilner}, D.~J., {Espaillat}, C., {et~al.}
  2011{\natexlab{b}}, \apj, 732, 42, \dodoi{10.1088/0004-637X/732/1/42}

\bibitem[{{Andrews} {et~al.}(2016){Andrews}, {Wilner}, {Zhu}, {Birnstiel},
  {Carpenter}, {P{\'e}rez}, {Bai}, {{\"O}berg}, {Hughes}, {Isella}, \&
  {Ricci}}]{andrews16}
{Andrews}, S.~M., {Wilner}, D.~J., {Zhu}, Z., {et~al.} 2016, \apjl, 820, L40,
  \dodoi{10.3847/2041-8205/820/2/L40}

\bibitem[{{Bai} \& {Stone}(2014)}]{bai14b}
{Bai}, X.-N., \& {Stone}, J.~M. 2014, \apj, 796, 31,
  \dodoi{10.1088/0004-637X/796/1/31}

\bibitem[{{Banzatti} {et~al.}(2015){Banzatti}, {Pinilla}, {Ricci},
  {Pontoppidan}, {Birnstiel}, \& {Ciesla}}]{banzatti15}
{Banzatti}, A., {Pinilla}, P., {Ricci}, L., {et~al.} 2015, \apjl, 815, L15,
  \dodoi{10.1088/2041-8205/815/1/L15}

\bibitem[{{Benisty} {et~al.}(2010){Benisty}, {Natta}, {Isella}, {Berger},
  {Massi}, {Le Bouquin}, {M{\'e}rand}, {Duvert}, {Kraus}, {Malbet}, {Olofsson},
  {Robbe-Dubois}, {Testi}, {Vannier}, \& {Weigelt}}]{benisty10}
{Benisty}, M., {Natta}, A., {Isella}, A., {et~al.} 2010, \aap, 511, A74,
  \dodoi{10.1051/0004-6361/200912898}

\bibitem[{{Benisty} {et~al.}(2017){Benisty}, {Stolker}, {Pohl}, {de Boer},
  {Lesur}, {Dominik}, {Dullemond}, {Langlois}, {Min}, {Wagner}, {Henning},
  {Juhasz}, {Pinilla}, {Facchini}, {Apai}, {van Boekel}, {Garufi}, {Ginski},
  {M{\'e}nard}, {Pinte}, {Quanz}, {Zurlo}, {Boccaletti}, {Bonnefoy}, {Beuzit},
  {Chauvin}, {Cudel}, {Desidera}, {Feldt}, {Fontanive}, {Gratton}, {Kasper},
  {Lagrange}, {LeCoroller}, {Mouillet}, {Mesa}, {Sissa}, {Vigan}, {Antichi},
  {Buey}, {Fusco}, {Gisler}, {Llored}, {Magnard}, {Moeller-Nilsson}, {Pragt},
  {Roelfsema}, {Sauvage}, \& {Wildi}}]{benisty17}
{Benisty}, M., {Stolker}, T., {Pohl}, A., {et~al.} 2017, \aap, 597, A42,
  \dodoi{10.1051/0004-6361/201629798}

\bibitem[{{B{\'e}thune} {et~al.}(2017){B{\'e}thune}, {Lesur}, \&
  {Ferreira}}]{bethune17}
{B{\'e}thune}, W., {Lesur}, G., \& {Ferreira}, J. 2017, \aap, 600, A75,
  \dodoi{10.1051/0004-6361/201630056}

\bibitem[{{Boehler} {et~al.}(2018){Boehler}, {Ricci}, {Weaver}, {Isella},
  {Benisty}, {Carpenter}, {Grady}, {Shen}, {Tang}, \& {Perez}}]{boehler18}
{Boehler}, Y., {Ricci}, L., {Weaver}, E., {et~al.} 2018, \apj, 853, 162,
  \dodoi{10.3847/1538-4357/aaa19c}

\bibitem[{{Bohren} \& {Huffman}(1998)}]{bohren98}
{Bohren}, C.~F., \& {Huffman}, D.~R. 1998, {Absorption and Scattering of Light
  by Small Particles}, 544

\bibitem[{{Bryden} {et~al.}(1999){Bryden}, {Chen}, {Lin}, {Nelson}, \&
  {Papaloizou}}]{bryden99}
{Bryden}, G., {Chen}, X., {Lin}, D.~N.~C., {Nelson}, R.~P., \& {Papaloizou},
  J.~C.~B. 1999, \apj, 514, 344, \dodoi{10.1086/306917}

\bibitem[{{Casassus} {et~al.}(2013){Casassus}, {van der Plas}, {M}, {Dent},
  {Fomalont}, {Hagelberg}, {Hales}, {Jord{\'a}n}, {Mawet}, {M{\'e}nard},
  {Wootten}, {Wilner}, {Hughes}, {Schreiber}, {Girard}, {Ercolano}, {Canovas},
  {Rom{\'a}n}, \& {Salinas}}]{casassus13}
{Casassus}, S., {van der Plas}, G., {M}, S.~P., {et~al.} 2013, \nat, 493, 191,
  \dodoi{10.1038/nature11769}

\bibitem[{{de Gregorio-Monsalvo} {et~al.}(2013){de Gregorio-Monsalvo},
  {M{\'e}nard}, {Dent}, {Pinte}, {L{\'o}pez}, {Klaassen}, {Hales},
  {Cort{\'e}s}, {Rawlings}, {Tachihara}, {Testi}, {Takahashi}, {Chapillon},
  {Mathews}, {Juhasz}, {Akiyama}, {Higuchi}, {Saito}, {Nyman}, {Phillips},
  {Rod{\'o}n}, {Corder}, \& {Van Kempen}}]{degregorio-monsalvo13}
{de Gregorio-Monsalvo}, I., {M{\'e}nard}, F., {Dent}, W., {et~al.} 2013, \aap,
  557, A133, \dodoi{10.1051/0004-6361/201321603}

\bibitem[{{Dong} {et~al.}(2012){Dong}, {Rafikov}, {Zhu}, {Hartmann}, {Whitney},
  {Brandt}, {Muto}, {Hashimoto}, {Grady}, {Follette}, {Kuzuhara}, {Tanii},
  {Itoh}, {Thalmann}, {Wisniewski}, {Mayama}, {Janson}, {Abe}, {Brandner},
  {Carson}, {Egner}, {Feldt}, {Goto}, {Guyon}, {Hayano}, {Hayashi}, {Hayashi},
  {Henning}, {Hodapp}, {Honda}, {Inutsuka}, {Ishii}, {Iye}, {Kandori}, {Knapp},
  {Kudo}, {Kusakabe}, {Matsuo}, {McElwain}, {Miyama}, {Morino}, {Moro-Martin},
  {Nishimura}, {Pyo}, {Suto}, {Suzuki}, {Takami}, {Takato}, {Terada}, {Tomono},
  {Turner}, {Watanabe}, {Yamada}, {Takami}, {Usuda}, \& {Tamura}}]{dong12}
{Dong}, R., {Rafikov}, R., {Zhu}, Z., {et~al.} 2012, \apj, 750, 161,
  \dodoi{10.1088/0004-637X/750/2/161}

\bibitem[{{Dong} {et~al.}(2017){Dong}, {van der Marel}, {Hashimoto}, {Chiang},
  {Akiyama}, {Liu}, {Muto}, {Knapp}, {Tsukagoshi}, {Brown}, {Bruderer},
  {Koyamatsu}, {Kudo}, {Ohashi}, {Rich}, {Satoshi}, {Takami}, {Wisniewski},
  {Yang}, {Zhu}, \& {Tamura}}]{dong17}
{Dong}, R., {van der Marel}, N., {Hashimoto}, J., {et~al.} 2017, \apj, 836,
  201, \dodoi{10.3847/1538-4357/aa5abf}

\bibitem[{{Dong} {et~al.}(2018){Dong}, {Liu}, {Eisner}, {Andrews}, {Fung},
  {Zhu}, {Chiang}, {Hashimoto}, {Liu}, {Casassus}, {Esposito}, {Hasegawa},
  {Muto}, {Pavlyuchenkov}, {Wilner}, {Akiyama}, {Tamura}, \&
  {Wisniewski}}]{dong18}
{Dong}, R., {Liu}, S.-y., {Eisner}, J., {et~al.} 2018, \apj, 860, 124,
  \dodoi{10.3847/1538-4357/aac6cb}

\bibitem[{{Fedele} {et~al.}(2017){Fedele}, {Carney}, {Hogerheijde}, {Walsh},
  {Miotello}, {Klaassen}, {Bruderer}, {Henning}, \& {van Dishoeck}}]{fedele17}
{Fedele}, D., {Carney}, M., {Hogerheijde}, M.~R., {et~al.} 2017, \aap, 600,
  A72, \dodoi{10.1051/0004-6361/201629860}

\bibitem[{{Fedele} {et~al.}(2018){Fedele}, {Tazzari}, {Booth}, {Testi},
  {Clarke}, {Pascucci}, {Kospal}, {Semenov}, {Bruderer}, {Henning}, \&
  {Teague}}]{fedele18}
{Fedele}, D., {Tazzari}, M., {Booth}, R., {et~al.} 2018, \aap, 610, A24,
  \dodoi{10.1051/0004-6361/201731978}

\bibitem[{{Flaherty} {et~al.}(2015){Flaherty}, {Hughes}, {Rosenfeld},
  {Andrews}, {Chiang}, {Simon}, {Kerzner}, \& {Wilner}}]{flaherty15}
{Flaherty}, K.~M., {Hughes}, A.~M., {Rosenfeld}, K.~A., {et~al.} 2015, \apj,
  813, 99, \dodoi{10.1088/0004-637X/813/2/99}

\bibitem[{{Flock} {et~al.}(2015){Flock}, {Ruge}, {Dzyurkevich}, {Henning},
  {Klahr}, \& {Wolf}}]{flock15}
{Flock}, M., {Ruge}, J.~P., {Dzyurkevich}, N., {et~al.} 2015, \aap, 574, A68,
  \dodoi{10.1051/0004-6361/201424693}

\bibitem[{{Foreman-Mackey} {et~al.}(2013){Foreman-Mackey}, {Hogg}, {Lang}, \&
  {Goodman}}]{foreman-mackey13}
{Foreman-Mackey}, D., {Hogg}, D.~W., {Lang}, D., \& {Goodman}, J. 2013, \pasp,
  125, 306, \dodoi{10.1086/670067}

\bibitem[{{Gaia Collaboration} {et~al.}(2018){Gaia Collaboration}, {Brown},
  {Vallenari}, {Prusti}, {de Bruijne}, {Babusiaux}, \&
  {Bailer-Jones}}]{gaia_dr2}
{Gaia Collaboration}, {Brown}, A.~G.~A., {Vallenari}, A., {et~al.} 2018, ArXiv
  e-prints.
\newblock \doarXiv{1804.09365}

\bibitem[{{Garufi} {et~al.}(2014){Garufi}, {Quanz}, {Schmid}, {Avenhaus},
  {Buenzli}, \& {Wolf}}]{garufi14}
{Garufi}, A., {Quanz}, S.~P., {Schmid}, H.~M., {et~al.} 2014, \aap, 568, A40,
  \dodoi{10.1051/0004-6361/201424262}

\bibitem[{{Goodman} \& {Weare}(2010)}]{goodman10}
{Goodman}, J., \& {Weare}, J. 2010, Comm.~App.~Math.~Comp.~Sci., 5, 65

\bibitem[{{Grady} {et~al.}(2000){Grady}, {Devine}, {Woodgate}, {Kimble},
  {Bruhweiler}, {Boggess}, {Linsky}, {Plait}, {Clampin}, \& {Kalas}}]{grady00}
{Grady}, C.~A., {Devine}, D., {Woodgate}, B., {et~al.} 2000, \apj, 544, 895,
  \dodoi{10.1086/317222}

\bibitem[{{Guidi} {et~al.}(2018){Guidi}, {Ruane}, {Williams}, {Mawet}, {Testi},
  {Zurlo}, {Absil}, {Bottom}, {Choquet}, {Christiaens}, {Femen{\'{\i}}a
  Castell{\'a}}, {Huby}, {Isella}, {Kastner}, {Meshkat}, {Reggiani}, {Riggs},
  {Serabyn}, \& {Wallack}}]{guidi18}
{Guidi}, G., {Ruane}, G., {Williams}, J.~P., {et~al.} 2018, \mnras, 479, 1505,
  \dodoi{10.1093/mnras/sty1642}

\bibitem[{{Isella} {et~al.}(2014){Isella}, {Chandler}, {Carpenter},
  {P{\'e}rez}, \& {Ricci}}]{isella14}
{Isella}, A., {Chandler}, C.~J., {Carpenter}, J.~M., {P{\'e}rez}, L.~M., \&
  {Ricci}, L. 2014, \apj, 788, 129, \dodoi{10.1088/0004-637X/788/2/129}

\bibitem[{{Isella} {et~al.}(2013){Isella}, {P{\'e}rez}, {Carpenter}, {Ricci},
  {Andrews}, \& {Rosenfeld}}]{isella13}
{Isella}, A., {P{\'e}rez}, L.~M., {Carpenter}, J.~M., {et~al.} 2013, \apj, 775,
  30, \dodoi{10.1088/0004-637X/775/1/30}

\bibitem[{{Isella} {et~al.}(2007){Isella}, {Testi}, {Natta}, {Neri}, {Wilner},
  \& {Qi}}]{isella07}
{Isella}, A., {Testi}, L., {Natta}, A., {et~al.} 2007, \aap, 469, 213,
  \dodoi{10.1051/0004-6361:20077385}

\bibitem[{{Isella} {et~al.}(2016){Isella}, {Guidi}, {Testi}, {Liu}, {Li}, {Li},
  {Weaver}, {Boehler}, {Carpenter}, {de Gregorio-Monsalvo}, {Manara}, {Natta},
  {P{\'e}rez}, {Ricci}, {Sargent}, {Tazzari}, \& {Turner}}]{isella16}
{Isella}, A., {Guidi}, G., {Testi}, L., {et~al.} 2016, \prl, 117, 251101

\bibitem[{{Isella} {et~al.}(2018){Isella}, {Andrews}, {Bai}, {Benisty},
  {Birnstiel}, {Carpenter}, {Dullemond}, {Guzm{\'a}n}, {Huang}, {Hughes},
  {{\"O}berg}, {P{\'e}rez}, {Ricci}, {Troncoso}, {Weaver}, {Wilner}, {Zhang},
  \& {Zhu}}]{isella18}
{Isella}, A., {Andrews}, S.~M., {Bai}, X., {et~al.} 2018, \apjl

\bibitem[{{Jin} {et~al.}(2016){Jin}, {Li}, {Isella}, {Li}, \& {Ji}}]{jin17}
{Jin}, S., {Li}, S., {Isella}, A., {Li}, H., \& {Ji}, J. 2016, \apj, 818, 76,
  \dodoi{10.3847/0004-637X/818/1/76}

\bibitem[{{Keppler} {et~al.}(2018){Keppler}, {Benisty}, {M{\"u}ller},
  {Henning}, {van Boekel}, {Cantalloube}, {Ginski}, {van Holstein}, {Maire},
  {Pohl}, {Samland}, {Avenhaus}, {Baudino}, {Boccaletti}, {de Boer},
  {Bonnefoy}, {Chauvin}, {Desidera}, {Langlois}, {Lazzoni}, {Marleau},
  {Mordasini}, {Pawellek}, {Stolker}, {Vigan}, {Zurlo}, {Birnstiel},
  {Brandner}, {Feldt}, {Flock}, {Girard}, {Gratton}, {Hagelberg}, {Isella},
  {Janson}, {Juhasz}, {Kemmer}, {Kral}, {Lagrange}, {Launhardt}, {Matter},
  {M{\'e}nard}, {Milli}, {Molli{\`e}re}, {Olofsson}, {Perez}, {Pinilla},
  {Pinte}, {Quanz}, {Schmidt}, {Udry}, {Wahhaj}, {Williams}, {Buenzli},
  {Cudel}, {Dominik}, {Galicher}, {Kasper}, {Lannier}, {Mesa}, {Mouillet},
  {Peretti}, {Perrot}, {Salter}, {Sissa}, {Wildi}, {Abe}, {Antichi},
  {Augereau}, {Baruffolo}, {Baudoz}, {Bazzon}, {Beuzit}, {Blanchard}, {Brems},
  {Buey}, {De Caprio}, {Carbillet}, {Carle}, {Cascone}, {Cheetham}, {Claudi},
  {Costille}, {Delboulb{\'e}}, {Dohlen}, {Fantinel}, {Feautrier}, {Fusco},
  {Giro}, {Gisler}, {Gluck}, {Gry}, {Hubin}, {Hugot}, {Jaquet}, {Le Mignant},
  {Llored}, {Madec}, {Magnard}, {Martinez}, {Maurel}, {Meyer},
  {Moeller-Nilsson}, {Moulin}, {Mugnier}, {Origne}, {Pavlov}, {Perret},
  {Petit}, {Pragt}, {Puget}, {Rabou}, {Ramos}, {Rigal}, {Rochat}, {Roelfsema},
  {Rousset}, {Roux}, {Salasnich}, {Sauvage}, {Sevin}, {Soenke}, {Stadler},
  {Suarez}, {Turatto}, \& {Weber}}]{keppler18}
{Keppler}, M., {Benisty}, M., {M{\"u}ller}, A., {et~al.} 2018, ArXiv e-prints.
\newblock \doarXiv{1806.11568}

\bibitem[{{Klaassen} {et~al.}(2013){Klaassen}, {Juhasz}, {Mathews}, {Mottram},
  {De Gregorio-Monsalvo}, {van Dishoeck}, {Takahashi}, {Akiyama}, {Chapillon},
  {Espada}, {Hales}, {Hogerheijde}, {Rawlings}, {Schmalzl}, \&
  {Testi}}]{klaassen13}
{Klaassen}, P.~D., {Juhasz}, A., {Mathews}, G.~S., {et~al.} 2013, \aap, 555,
  A73, \dodoi{10.1051/0004-6361/201321129}

\bibitem[{Li {et~al.}(2001)Li, Colgate, Wendroff, \& Liska}]{li2001}
Li, H., Colgate, S.~A., Wendroff, B., \& Liska, R. 2001, The Astrophysical
  Journal, 551, 874

\bibitem[{{Liu} {et~al.}(2018){Liu}, {Jin}, {Li}, {Isella}, \& {Li}}]{liu18}
{Liu}, S.-F., {Jin}, S., {Li}, S., {Isella}, A., \& {Li}, H. 2018, \apj, 857,
  87, \dodoi{10.3847/1538-4357/aab718}

\bibitem[{{Marino} {et~al.}(2015){Marino}, {Perez}, \& {Casassus}}]{marino15}
{Marino}, S., {Perez}, S., \& {Casassus}, S. 2015, \apjl, 798, L44,
  \dodoi{10.1088/2041-8205/798/2/L44}

\bibitem[{{Mendigut{\'{\i}}a} {et~al.}(2018){Mendigut{\'{\i}}a}, {Oudmaijer},
  {Schneider}, {Hu{\'e}lamo}, {Baines}, {Brittain}, \&
  {Aberasturi}}]{mendigutia18}
{Mendigut{\'{\i}}a}, I., {Oudmaijer}, R.~D., {Schneider}, P.~C., {et~al.} 2018,
  \aap, 618, L9, \dodoi{10.1051/0004-6361/201834233}

\bibitem[{{Miranda} {et~al.}(2017){Miranda}, {Li}, {Li}, \& {Jin}}]{miranda17}
{Miranda}, R., {Li}, H., {Li}, S., \& {Jin}, S. 2017, \apj, 835, 118,
  \dodoi{10.3847/1538-4357/835/2/118}

\bibitem[{{Muro-Arena} {et~al.}(2018){Muro-Arena}, {Dominik}, {Waters}, {Min},
  {Klarmann}, {Ginski}, {Isella}, {Benisty}, {Pohl}, {Garufi}, {Hagelberg},
  {Langlois}, {Menard}, {Pinte}, {Sezestre}, {van der Plas}, {Villenave},
  {Delboulb{\'e}}, {Magnard}, {M{\"o}ller-Nilsson}, {Pragt}, {Rabou}, \&
  {Roelfsema}}]{muro18}
{Muro-Arena}, G.~A., {Dominik}, C., {Waters}, L.~B.~F.~M., {et~al.} 2018, \aap,
  614, A24, \dodoi{10.1051/0004-6361/201732299}

\bibitem[{{Okuzumi} {et~al.}(2016){Okuzumi}, {Momose}, {Sirono}, {Kobayashi},
  \& {Tanaka}}]{okuzumi16}
{Okuzumi}, S., {Momose}, M., {Sirono}, S.-i., {Kobayashi}, H., \& {Tanaka}, H.
  2016, \apj, 821, 82, \dodoi{10.3847/0004-637X/821/2/82}

\bibitem[{{Pinte} {et~al.}(2018){Pinte}, {M{\'e}nard}, {Duch{\^e}ne}, {Hill},
  {Dent}, {Woitke}, {Maret}, {van der Plas}, {Hales}, {Kamp}, {Thi}, {de
  Gregorio-Monsalvo}, {Rab}, {Quanz}, {Avenhaus}, {Carmona}, \&
  {Casassus}}]{pinte18}
{Pinte}, C., {M{\'e}nard}, F., {Duch{\^e}ne}, G., {et~al.} 2018, \aap, 609,
  A47, \dodoi{10.1051/0004-6361/201731377}

\bibitem[{{Rosenfeld} {et~al.}(2013){Rosenfeld}, {Andrews}, {Hughes}, {Wilner},
  \& {Qi}}]{rosenfeld13a}
{Rosenfeld}, K.~A., {Andrews}, S.~M., {Hughes}, A.~M., {Wilner}, D.~J., \&
  {Qi}, C. 2013, \apj, 774, 16, \dodoi{10.1088/0004-637X/774/1/16}

\bibitem[{{Sallum} {et~al.}(2015){Sallum}, {Follette}, {Eisner}, {Close},
  {Hinz}, {Kratter}, {Males}, {Skemer}, {Macintosh}, {Tuthill}, {Bailey},
  {Defr{\`e}re}, {Morzinski}, {Rodigas}, {Spalding}, {Vaz}, \&
  {Weinberger}}]{sallum15}
{Sallum}, S., {Follette}, K.~B., {Eisner}, J.~A., {et~al.} 2015, \nat, 527,
  342, \dodoi{10.1038/nature15761}

\bibitem[{{Suriano} {et~al.}(2018){Suriano}, {Li}, {Krasnopolsky}, \&
  {Shang}}]{suriano18}
{Suriano}, S.~S., {Li}, Z.-Y., {Krasnopolsky}, R., \& {Shang}, H. 2018, \mnras,
  477, 1239, \dodoi{10.1093/mnras/sty717}

\bibitem[{{Tang} {et~al.}(2017){Tang}, {Guilloteau}, {Dutrey}, {Muto}, {Shen},
  {Gu}, {Inutsuka}, {Momose}, {Pietu}, {Fukagawa}, {Chapillon}, {Ho}, {di
  Folco}, {Corder}, {Ohashi}, \& {Hashimoto}}]{tang17}
{Tang}, Y.-W., {Guilloteau}, S., {Dutrey}, A., {et~al.} 2017, \apj, 840, 32,
  \dodoi{10.3847/1538-4357/aa6af7}

\bibitem[{{Tazzari} {et~al.}(2017){Tazzari}, {Testi}, {Natta}, {Ansdell},
  {Carpenter}, {Guidi}, {Hogerheijde}, {Manara}, {Miotello}, {van der Marel},
  {van Dishoeck}, \& {Williams}}]{tazzari17}
{Tazzari}, M., {Testi}, L., {Natta}, A., {et~al.} 2017, ArXiv e-prints.
\newblock \doarXiv{1707.01499}

\bibitem[{{Teague} {et~al.}(2018){Teague}, {Bae}, {Bergin}, {Birnstiel}, \&
  {Foreman-Mackey}}]{teague18}
{Teague}, R., {Bae}, J., {Bergin}, E.~A., {Birnstiel}, T., \& {Foreman-Mackey},
  D. 2018, \apjl, 860, L12, \dodoi{10.3847/2041-8213/aac6d7}

\bibitem[{{Thalmann} {et~al.}(2016){Thalmann}, {Janson}, {Garufi},
  {Boccaletti}, {Quanz}, {Sissa}, {Gratton}, {Salter}, {Benisty}, {Bonnefoy},
  {Chauvin}, {Daemgen}, {Desidera}, {Dominik}, {Engler}, {Feldt}, {Henning},
  {Lagrange}, {Langlois}, {Lannier}, {Le Coroller}, {Ligi}, {M{\'e}nard},
  {Mesa}, {Meyer}, {Mulders}, {Olofsson}, {Pinte}, {Schmid}, {Vigan}, \&
  {Zurlo}}]{thalmann16}
{Thalmann}, C., {Janson}, M., {Garufi}, A., {et~al.} 2016, \apjl, 828, L17,
  \dodoi{10.3847/2041-8205/828/2/L17}

\bibitem[{{Tilling} {et~al.}(2012){Tilling}, {Woitke}, {Meeus}, {Mora},
  {Montesinos}, {Riviere-Marichalar}, {Eiroa}, {Thi}, {Isella}, {Roberge},
  {Martin-Zaidi}, {Kamp}, {Pinte}, {Sandell}, {Vacca}, {M{\'e}nard},
  {Mendigut{\'{\i}}a}, {Duch{\^e}ne}, {Dent}, {Aresu}, {Meijerink}, \&
  {Spaans}}]{tilling12}
{Tilling}, I., {Woitke}, P., {Meeus}, G., {et~al.} 2012, \aap, 538, A20,
  \dodoi{10.1051/0004-6361/201116919}

\bibitem[{{van der Marel} {et~al.}(2016){van der Marel}, {Cazzoletti},
  {Pinilla}, \& {Garufi}}]{vandermarel16}
{van der Marel}, N., {Cazzoletti}, P., {Pinilla}, P., \& {Garufi}, A. 2016,
  \apj, 832, 178, \dodoi{10.3847/0004-637X/832/2/178}

\bibitem[{{van der Marel} {et~al.}(2015){van der Marel}, {van Dishoeck},
  {Bruderer}, {Perez}, \& {Isella}}]{vandermarel15}
{van der Marel}, N., {van Dishoeck}, E., {Bruderer}, S., {Perez}, L.~M., \&
  {Isella}, A. 2015, ArXiv e-prints.
\newblock \doarXiv{1504.03927}

\bibitem[{{van der Marel} {et~al.}(2018{\natexlab{a}}){van der Marel},
  {Williams}, \& {Bruderer}}]{vandermarel18b}
{van der Marel}, N., {Williams}, J.~P., \& {Bruderer}, S. 2018{\natexlab{a}},
  \apjl, 867, L14, \dodoi{10.3847/2041-8213/aae88e}

\bibitem[{{van der Marel} {et~al.}(2013){van der Marel}, {van Dishoeck},
  {Bruderer}, {Birnstiel}, {Pinilla}, {Dullemond}, {van Kempen}, {Schmalzl},
  {Brown}, {Herczeg}, {Mathews}, \& {Geers}}]{vandermarel13}
{van der Marel}, N., {van Dishoeck}, E.~F., {Bruderer}, S., {et~al.} 2013,
  Science, 340, 1199, \dodoi{10.1126/science.1236770}

\bibitem[{{van der Marel} {et~al.}(2018{\natexlab{b}}){van der Marel},
  {Williams}, {Ansdell}, {Manara}, {Miotello}, {Tazzari}, {Testi},
  {Hogerheijde}, {Bruderer}, {van Terwisga}, \& {van Dishoeck}}]{vandermarel18}
{van der Marel}, N., {Williams}, J.~P., {Ansdell}, M., {et~al.}
  2018{\natexlab{b}}, \apj, 854, 177, \dodoi{10.3847/1538-4357/aaaa6b}

\bibitem[{{Weaver} {et~al.}(2018){Weaver}, {Isella}, \& {Boehler}}]{weaver18}
{Weaver}, E., {Isella}, A., \& {Boehler}, Y. 2018, \apj, 853, 113,
  \dodoi{10.3847/1538-4357/aaa481}

\bibitem[{{Wisniewski} {et~al.}(2008){Wisniewski}, {Clampin}, {Grady},
  {Ardila}, {Ford}, {Golimowski}, {Illingworth}, \& {Krist}}]{wisniewski08}
{Wisniewski}, J.~P., {Clampin}, M., {Grady}, C.~A., {et~al.} 2008, \apj, 682,
  548, \dodoi{10.1086/589629}

\bibitem[{{Zhang} {et~al.}(2016){Zhang}, {Bergin}, {Blake}, {Cleeves},
  {Hogerheijde}, {Salinas}, \& {Schwarz}}]{zhang16}
{Zhang}, K., {Bergin}, E.~A., {Blake}, G.~A., {et~al.} 2016, \apjl, 818, L16,
  \dodoi{10.3847/2041-8205/818/1/L16}

\bibitem[{{Zhang} {et~al.}(2015){Zhang}, {Blake}, \& {Bergin}}]{zhang15}
{Zhang}, K., {Blake}, G.~A., \& {Bergin}, E.~A. 2015, \apjl, 806, L7,
  \dodoi{10.1088/2041-8205/806/1/L7}

\bibitem[{{Zhang} {et~al.}(2014){Zhang}, {Isella}, {Carpenter}, \&
  {Blake}}]{zhang14}
{Zhang}, K., {Isella}, A., {Carpenter}, J.~M., \& {Blake}, G.~A. 2014, \apj,
  791, 42, \dodoi{10.1088/0004-637X/791/1/42}

\bibitem[{{Zhu} {et~al.}(2014){Zhu}, {Stone}, {Rafikov}, \& {Bai}}]{zhu14}
{Zhu}, Z., {Stone}, J.~M., {Rafikov}, R.~R., \& {Bai}, X.-n. 2014, \apj, 785,
  122, \dodoi{10.1088/0004-637X/785/2/122}

\end{thebibliography}



\end{document}